\documentclass[%
reprint,
superscriptaddress,
showpacs,
hidelinks,
amsmath,amssymb,
aps,
prb,
a4paper
]{revtex4-1}
\usepackage{graphicx,epstopdf}
\usepackage{placeins}
\usepackage{tikz}
\usepackage{tikz-3dplot}
\usetikzlibrary{calc,math}
\usetikzlibrary{intersections}
\usetikzlibrary{positioning}
\usetikzlibrary{arrows,decorations.markings}
\usepackage{pgfplots}
\pgfplotsset{compat=newest}
\usepgflibrary{shapes.geometric}
\usepackage{mathrsfs}
\usepackage{bm}
\usepackage{esint}
\usepackage{siunitx}
\usepackage{physics}
\usepackage{xurl}

\usepackage[normalem]{ulem}
\usepackage{cancel}

\begin{document}
\title{Electron-hole coherence in core-shell nanowires \\ with partial proximity-induced superconductivity}
\author{Kristjan Ottar Klausen}
\affiliation{Department of Engineering, Reykjavik University, Menntavegur 1, IS-101 Reykjavik, Iceland.}
\author{Anna Sitek}
\affiliation{Department of Theoretical Physics,
	Wroclaw University of Science and Technology,
	Wybrze{\.z}e Wyspia{\'n}skiego 27, 50-370 Wroclaw, Poland.}
\author{Sigurdur I.\ Erlingsson}
\affiliation{Department of Engineering, Reykjavik University, Menntavegur 1, IS-101 Reykjavik, Iceland.}
\author{Andrei Manolescu}
\affiliation{Department of Engineering, Reykjavik University, Menntavegur 1, IS-101 Reykjavik, Iceland.}

\vspace{10 mm}

\begin{abstract}
By solving the Bogoliubov-de Gennes Hamiltonian, the electron-hole coherence within a partially proximitized n-doped semiconductor shell of a core-shell nanowire heterostructure is investigated numerically and compared with the Andreev reflection interpretation of proximity induced superconductivity. Partial proximitization is considered to quantify the effects of a reduced coherence length. Three cases of partial proximitization of the shell are explored: radial, angular and longitudinal. For the radial case, it is found that the boundary conditions impose localization probability maxima in the center of the shell in spite of off-center radial proximitization. The induced superconductivity gap is calculated
as a function of the ratio between the proximitized shell thickness and the total shell thickness.
In the angular case, the lowest energy state of a hexagonal wire with a single proximitized side is found to display the essence of Andreev reflection, only by lengthwise summation of the localization probability.
 In the longitudinal case, a clear correspondence with Andreev reflection is seen in the localization probability as a function of length along a half-proximitized wire. The effect of an external magnetic field oriented along the wire is explored.

\end{abstract}
\vspace{-20 mm}
\maketitle
\section{INTRODUCTION}
Semiconductor nanowires with proximity induced superconductivity have emerged as key elements in various platforms proposed to realize qubits and other emerging technologies at the quantum scale \cite{Hays_21,Aguado2020,Benito2020}.
The proximity effect is generally hypothesized to stem from electron-hole coherence, brought on by Andreev reflection at the superconductor interface, where an incoming electron is retro-reflected as a hole \cite{Andreev,AndreevProx,BTK}.
	The superconducting proximity effect has resurfaced time after time in the past decades as a hot research topic due to relevance to research topics in each decade \cite{TSC_prox,Tudor2014,Graphene_2013,Zha_2010,Klapwijk2004,JACOBS1999669,osti_5167845,Haesendonck_1981,Meissner71,clarke:jpa-00213516}, most recently due to the search for Majorana zero modes in nanostructures \cite{Elsa2020,prox2019_junger,Review_Agauado}. 
These zero modes are expected to be hosted in synthetic
topological superconductors, where p-wave superconductivity can be engineered using spin-orbit coupling in conjunction with Zeeman splitting and proximity induced superconductivity in semiconductors \cite{TudorBok,Alicea_2010,Sau2010}.

Core-shell nanowires are radial heterojunctions consisting of a core which is wrapped by one or more layers of different materials. Due to crystallographic structure they usually have polygonal cross-sections \cite{Blomers13,Rieger12,Haas13,Funk13,Erhard15,Weiss14b,Jadczak14,Qian04,Qian05,Baird09,Heurlin15,Dong09,Yuan15,Goransson19,Rieger15}, and thus the shells become prismatic nanotubes, but circular systems have also been obtained \cite{Kim2017}. The sharp corners of the cross-section induce non-uniform electron localization along the circumference of the tube, in particular, low energy electrons are accumulated in the vicinity of sharp edges, while carriers of higher energy are shifted to the facets \cite{Ferrari09b,Wong11,Sitek15,Sitek16}. If the shell is very thin then the low-energy electrons are depleted from the facets and the shell becomes a multiple-channel system consisting of well-separated 1D electron channels situated along the edges. Due to their unique localization and a variety of other interesting properties, core-shell nanowires have
 been extensively investigated in the last two decades \cite{Royo_2017,Shen2009}, showing promise in multiple applications such as lasers \cite{Koblm_ller_2017}, energy harvesting devices\cite{Florica2019,Hassan:19} and photovoltaics \cite{core-shell_photo}. 
By n-doping, the chemical potential can be moved into the conduction band such that electrons become the only charge carriers and the material behaves effectively as a metal with the effective mass of the host semiconductor.
Earlier investigations have indicated that 
due to 1D electron channels along the sharp edges of prismatic tubes, 
multiple Majorana Zero Modes can be hosted in a single core-shell nanowire \cite{Andrei,Ottar_Klausen_2020}.
However only if the electron-hole coherence length is larger than the whole structure, can the shell be considered fully proximitized and electron-hole coherence can be expected to be uniform.

In this paper, electron-hole coherence is investigated in an n-doped semiconductor core-shell nanowire with partial proximity induced superconductivity. 
Electron-hole coherence of the lowest energy states is compared with the Andreev reflection picture of proximitized superconductivity in the radial, angular and longitudinal interfaces arising within a single nanowire. 

\section{ELECTRON-HOLE COHERENCE AND THE PROXIMITY EFFECT}

One of the earlier theoretical descriptions of the spatial dependence of the order parameter in the superconducting proximity effect was done by McMillan\cite{MacMillan_NS_Theory}, using a Green's function approach based on the Gor'kov equations \cite{Gorkov_58} to describe a normal metal-superconductor (NS) junction. In this method, the BCS potential for a quasi 1D problem is written in terms of the pairing interaction $V(\bm{x})$ and the condensation amplitude\cite{Klapwijk2004} $F(\bm{x})$,
\begin{equation}
\Delta (\mathbf{x})=V(\mathbf{x})F(\mathbf{x}),
\end{equation}
where the ensemble average of field operators with opposite spin projections describes the condensation amplitude\cite{FetterW}
\begin{equation}
F(\mathbf{x}) =\langle \hat{\psi}_\uparrow(\mathbf{x})\hat{\psi}_\downarrow(\mathbf{x}) \rangle.
\end{equation}
McMillan \cite{MacMillan_NS_Theory} called the problem of the NS-junction possibly the simplest one in space-dependent superconductivity and proposed, in his own words, \lq\lq a very nearly complete solution\rq\rq \text{} of the problem for the case of infinite length of both metals, evaluating
\begin{equation}
F(\mathbf{x})= \frac{1}{\pi} \int_{0}^{E_{c_0}} \text{Im}[\mathcal{G}_{12}(E,\bm{x},\bm{x'})] \, dE ,
\end{equation}
where $E_{c_0}$ is the cut-off energy\cite{Zheng_2005}. $\mathcal{G}_{12}$ is the upper off-diagonal component of the $2\times2$ Green's function
\begin{equation}
	\mathcal{G}(\mathbf{x},t,\mathbf{x}',t') =-i\langle0| T\{\Psi(\mathbf{x},t) \Psi^\dagger(\mathbf{x}',t')\}|0\rangle,
	\label{GF_2x2}
	\end{equation}
written in the Nambu\cite{Nambu_1960} spinor formalism, 
\begin{equation}
\Psi (\mathbf{x})=
\begin{pmatrix}
\hat{\psi}_\uparrow(\mathbf{x}) \\ 
\hat{\psi}_\downarrow^\dagger(\mathbf{x}) 
\end{pmatrix},
\end{equation}
where $T$ denotes time ordering and $|0\rangle$ the Heisenberg ground state.
$F(\bm{x})$ is also known as the anomalous Green's function \cite{BruusFlensberg}.

\begin{figure}[ht!]
	\centering
     \includegraphics[width=0.48\textwidth]{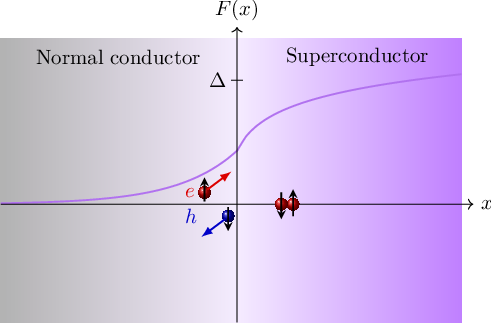}
	\vspace{-0.5cm}
	\caption{Simplified sketch of the proximity effect in terms of the condensation amplitude $F(x)$ and Andreev reflection. An electron with energy $E<\Delta$ at an N-S boundary will be \textit{retro-reflected} as a hole while forming a Cooper pair within the superconductor\cite{BTK}.}
	\label{AndreevRefl}
\end{figure}

Another fundamental reference in the field is a book chapter written by Deutscher and de Gennes \cite{DeGennesDeutcher} where the distinction between a clean and dirty junction is made and the following simplified result presented for the spatial dependence of $F(\mathbf{x})$. For a clean metal, where the mean free path $l_N$ therein is larger compared to the coherence length, $l_N > \xi_N$, the asymptotic form is obtained
\begin{equation}
F(\mathbf{x})=\phi(\mathbf{x})\exp(- \frac{2\pi k_B T}{\hbar v_F}\abs{\bm{x}}),
\end{equation}
where $\phi(\mathbf{x})$ is some slowly varying function, $k_B$ is the Boltzmann constant, $T$ denotes temperature and $v_F$ is the Fermi velocity.
For the limiting case of the temperature being close to zero a result by Falk\cite{Falk} is cited, 
\begin{equation}
F(\mathbf{x}) \sim \frac{1}{\abs{\bm{x}}} \ .
\end{equation}
Falks paper \cite{Falk} has a similar Green's function based approach to the Gor'kov equations as McMillan \cite{MacMillan_NS_Theory} and proceeds McMillan's work by five years. Andreev\cite{Andreev} considered the equations of motion obeyed by components of the Green's function in Eq.\ \eqref{GF_2x2} in which diagonal and off-diagonals are coupled, to derive a curious scattering process at the plane boundary between the normal and superconducting phases which now bears his name.
Andreev reflection, Fig.\ \ref{AndreevRefl}, is the conjugate retro-reflection of electrons and holes at a metal-superconductor boundary \cite{Andreev}. Retro-reflection means that an incoming electron from the normal metal side is reflected such that it traces back the incident trajectory. 
	In order for an incident electron at the normal metal side with energy below the gap parameter $\Delta$, to be transferred across the boundary, the formation of a Cooper pair in the superconductor requires another electron with equal and opposite momentum, which can be seen as a reflected hole. 
\FloatBarrier
At present time, electron-hole coherence via Andreev reflection is considered to be the mechanism behind the superconducting proximity effect \cite{Tinkham, Schapers}. 
Blonder, Tinkham and Klapwijk (BTK) refined the scattering approach to the problem using the Bogoliubov equations and further computed I-V curves along with transmission and reflection coefficients for all cases of energy relative to the superconducting gap including a delta-barrier at the interface \cite{BTK}. This work has since become seminal for Andreev reflection and is fundamental to most tunneling spectroscopy experiments on superconducting junctions \cite{Tinkham}. The scattering formalism has the advantage of being readily interpreted and familiar from the standard educational problem in quantum mechanics of scattering from a potential barrier. 
Klapwijk \cite{Klapwijk2004} later noted that the following self-consistency equation was not incorporated in the original BTK approach, and the superconducting gap implemented as a step function at the interface. The self-consistency equation can be written as
\begin{equation}
\Delta(\mathbf{r})=V(\mathbf{r})F(\mathbf{r}) = V(\mathbf{r}) \sum_E u(\mathbf{r}) v^\dagger(\mathbf{r}) [1-2f(E)],
\end{equation}
where $u(\mathbf{r})$ and $v(\mathbf{r})$ are the electron- and hole components of the quasiparticle wavefunction respectively, and $f(E)$ is the Fermi-distribution function,
\begin{equation}
f(E)=[1+ \exp(E-\mu/(k_B T))]^{-1} \ .
\end{equation}
Even if the pairing interaction $V(\mathbf{r})$ is zero in the normal metal, $F(\mathbf{r})$ can be non-zero, stemming from electron-hole coherence, which can be interpreted as the superconductivity leakage in the normal metal \cite{Klapwijk2004}. The self-consistency equation determines the variation of $\Delta(\mathbf{r})$ at the intersection but the general features of Andreev reflection are independent of it \cite{Beenakker91}. Self-consistency has been shown to be of great importance for interfaces with d-wave superconductivity \cite{MARTIN99,Aizawa_2018}.
The spatial propagation of the superconducting order parameter at NS interfaces has been found \cite{Devoret_96}  to be adequately described by the Usadel equations \cite{Usadel_70}, which are a diffusive (dirty) limit of the Eilenberger-Larkin-Ovchinnikov equations \cite{Eilenberger_68,LarkinOvchinnikov_69}, which are a quasiclassical approximation to the Gorkov equations \cite{Gorkov_58}. The Gorkov equations can be used to derive the semi-classical Ginzburg-Landau theory \cite{GL} from a microscopic description \cite{Gorkov_59}.
	
Considerable work has been done in the past decade on the many subtleties of the superconducting gap parameter in hybrid semiconductor-superconductor systems in relation to the quest for experimental realization of Majorana Zero Modes \cite{Ridderbos2020,Reeg_2017,Cole15,Loss_12,Osca_delta},
in particular, where p-wave superconductivity is engineered in nanowires by combining the proximity effect, Zeeman and Rashba spin-orbit interaction \cite{Alicea_2010}. Another promising approach are hybrid systems using ferromagnetic materials \cite{Livanas_2021} to obtain spin-triplet pairing\cite{Klaus_2015,Zha_2015,Klaus_2012, Klaus_2008, Klaus_2007,Fominov_2003} allowing for topological states \cite{Hui_2015}. Properties of the induced gap can be significantly altered in the presence of ferromagnetism  \cite{Klaus_2018_SF,Klaus_2005,Klaus_2003,Bergeret_2001,Klaus_2001}.
Furthermore, interfaces of superconductors and ferromagnetic materials have opened up possibilities for superconducting spintronics \cite{SCspintronics_21,Khusainov_02}, allowing for spin-dependent control of supercurrents \cite{Klaus_2016,SpinSwitch} and transition temperatures \cite{Klaus_2018_spinvalve,Budzin_05}.

\section{MODEL AND METHODS}
A three dimensional proximitized core-shell nanowire is modeled in the zero temperature limit using cylindrical coordinates, where the $z$-axis is defined along the wire growth direction. Using a divide and conquer algorithm\cite{lapack99} coded in Fortran\cite{FastFortran}, energy spectra and states of the system are obtained by numerical diagonalization of the the Bogoliubov-de Gennes (BdG) Hamiltonian \cite{Jianxin, Bogoliubov:1958km},
\begin{equation}
	H_{BdG}= 
	\begin{pmatrix}
		[H_w - \mu]&\pm\Delta_{\uparrow\downarrow}\\
		\mp \Delta_{\uparrow\downarrow}^*&-[H^*_{w}-\mu]
	\end{pmatrix}.
	\label{BdGHam2x2}
\end{equation}
The matrix elements are written in the composite basis $|q \rangle$ consisting of the transverse modes $|a \rangle$, longitudinal modes $|n \rangle$, spin $|\sigma \rangle$ and particle-hole eigenstates $|\eta \rangle$ such that
\begin{equation}
|q\rangle =|\eta a n \sigma \rangle,
\end{equation}
where $|an \sigma \rangle$ are the eigenstates of the Hamiltonian for the wire without proximity induced superconductivity, 
\begin{equation}
H_\mathrm{w}= H_\mathrm{t} + H_\mathrm{l} + H_\mathrm{Z} \ .
\label{wireHamiltonian}
\end{equation}
The transverse and longitudinal components of the Hamiltonian are written as
\begin{equation}
H_\mathrm{t}+H_\mathrm{l}= \frac{(p_{\phi}+eA_{\phi})^2}{2m_e} 
-\frac{\hbar^2}{2m_er} \frac{\partial}{\partial r} 
\left(r\frac{\partial}{\partial r}\right)\ + \frac{p_z^2}{2m_e},
\end{equation}
where $A_{\phi}=\frac{1}{2}Br$ is the vector potential in the symmetric gauge, incorporating an external magnetic field, $\mathbf{B}$, directed along the wire axis.
The transverse eigenstates are expanded in terms of the lattice sites
\begin{equation}
|a \rangle = \sum_\kappa c_a |r_\kappa \phi_\kappa\rangle ,
\end{equation}
and matrix elements obtained by finite-difference discretization of derivatives \cite{Sitek15,Sitek16}. The cross-section geometry is added by infinite potential boundary conditions defining the hexagonal shape. For an infinite wire, the longitudinal modes are expressed in an exponential plane wave basis so that the Hamiltonian becomes a function of the longitudinal wave vector. For a finite wire, they are expanded in a sine basis,
\begin{equation}
|n \rangle =L_z^{-1/2}\sqrt{2} \sin\left(n\pi \left(\frac{z}{L_z}+\frac{1}{2}\right)\right) \ .
\label{zbasis}
\end{equation}
The length of the wire is $L_z$, the origin is defined in the nanowire center so that the wire spans the interval $\left[\frac{-L_z}{2}, \frac{L_z}{2}\right]$ along the $z$ axis.
The external magnetic field $B$ gives rise to the Zeeman term
\begin{equation}
H_\mathrm{Z} = -g^* \mu_B \sigma B,
\end{equation}
where $g^*$ is the effective Land\'{e} g-factor and $\mu_B$ the Bohr magneton.
The particle-hole symmetry and coupling are contained in the quantum number $\eta=\pm1$ and the matrix elements of the BdG Hamiltonian are obtained by the following, for $\eta=\eta'$

\begin{align}
\begin{split}
&\langle an \sigma  \eta |H_\mathrm{BdG}|a'n' \sigma'\eta' \rangle=
\eta[\text{Re}\langle an \sigma |H_\mathrm{w}|a'n' \sigma' \rangle \\&+ i\eta \langle an \sigma|H_\mathrm{w}|a'n' \sigma'\rangle - \mu \delta_{(an \sigma) (a'n' \sigma')}],
\end{split}
\label{Diag}
\end{align}
and for $\eta \neq \eta'$,
\begin{equation}
\langle an \sigma \eta |H_\mathrm{BdG}|a'n' \sigma'\eta' \rangle= \eta \sigma \delta_{\sigma,-\sigma'} \delta_{aa'} \delta_{nn'} \Delta_s \ .
\end{equation}
The chemical potential, $\mu$, is set to correspond to an n-doped semiconductor such that electrons are the main carriers of the system.
In accordance with the original works of Andreev \cite{Andreev} and BTK \cite{BTK}, the induced gap is a fixed parameter and spin-orbit coupling is not considered.
Partial proximitization is implemented in the superconducting gap parameter, $\Delta_s(r,\phi,z)$, by step functions of position in the radial, angular and longitudinal direction of the shell, so that ideal junctions with no interface barriers \cite{Reeg_2016} are formed in each case. No further boundary condition is imposed at the interface apart from the step-function of $\Delta_s$, which is known as a rigid boundary condition \cite{Beenakker_2005}.
Model parameters are set to correspond to an InSb shell with $\gamma=\frac{1}{2} g^* m_e =0.393$, and $\Delta= 0.50 \text{ meV}$. 
The numerical simulations were performed for single shell nanowires with cross-section diameter of 100 nm and shell thickness $d=10$ nm, the length was set to 10 $\mu$m in sections \ref{rad} and \ref{ang} along with a shorter wire of 1  $\mu$m and infinite one explored in sections \ref{long} and \ref{rad}, respectively.

\section{PARTIAL RADIAL PROXIMITIZATION}
\label{rad}
\begin{figure}[ht!]
	\centering
	\includegraphics[width=0.46\textwidth]{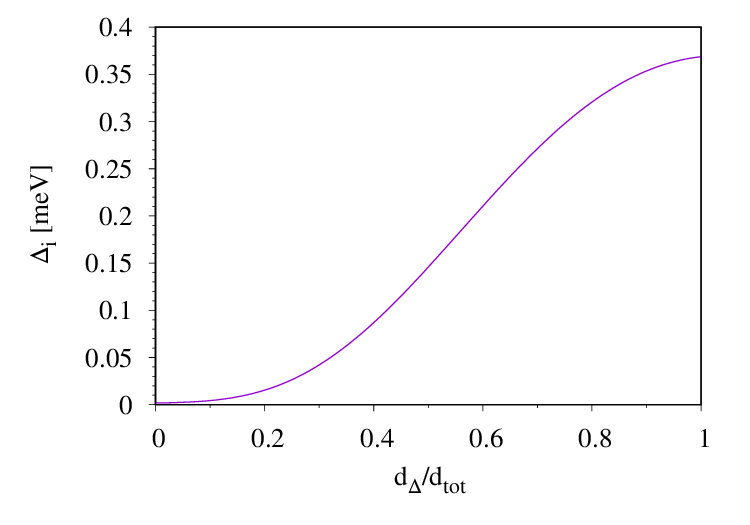}
	\caption{Induced superconducting energy gap of a partially proximitized nanowire as a function of the ratio of the proximitized shell thickness, $d_\Delta$, to the total shell thickness, $d_{tot}$.}
	\label{deltaar}
\end{figure}

 A core-shell nanowire fully coated with a superconducting layer 	
has a radially symmetric proximity induced gap in the shell. 
\begin{figure}[b!]
	\centering
	\includegraphics[width=0.49\textwidth]{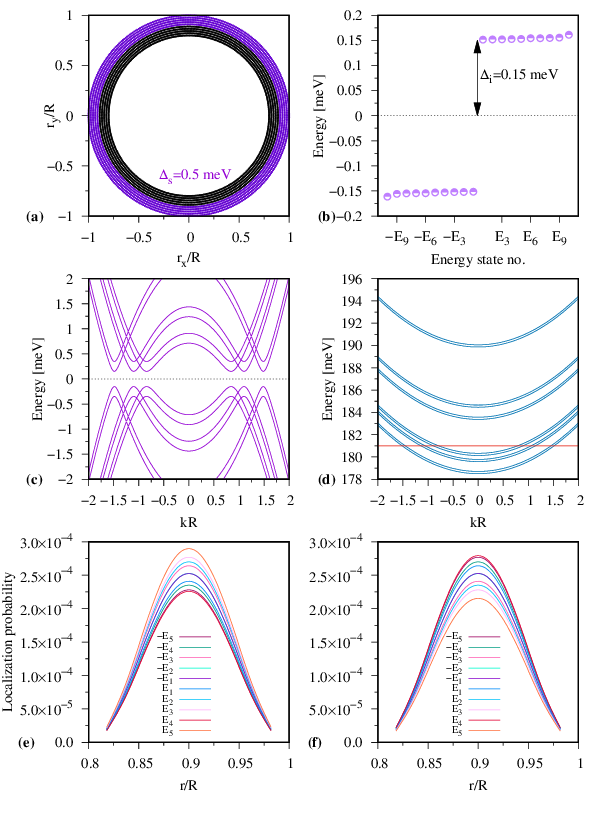}
	\caption{(a) Partial radial proximitization of the nanowire shell, diameter of $2R$. Half-proximitized shell with $\Delta_s=0.5 \text{ meV}$ in the outer half (purple) of the shell. (b) Finite wire BdG spectrum of the nanowire system, showing the induced gap $\Delta_i=0.15 \text{ meV}$. (c) Infinite wire BdG energy dispersion in terms of the dimensionless product of the wavevector, $k$, and radius, R. (d) Energy dispersion and chemical potential (red) of the infinite wire. (e) Longitudinal summation of localization probability on interior sites for the hole component $|v|^2$ of the lowest positive and negative energy states, for a single angular slice. (f) Corresponding electronic component $|u|^2$.}
	\label{3x2r}
\end{figure}
 Full-shell nanowire systems allow for additional control of the superconducting energy gap due to the Little-Parks effect \cite{FluxInducedMajorana2018,evenodd,Marcus21}. 
Multiple fabrication-specific microscopic details and material properties can influence the strength of the induced gap \cite{Stanescu_2022}. To investigate electron-hole coherence within the partially proximitized shell, the semiconductor shell is considered to have a superconducting gap $\Delta_s$ in its outer half, brought on by a surrounding superconductor, not included in the Hamiltonian. A cylindrical shell is studied, to isolate the effect of a partial proximitization radially, Fig.\ \ref{3x2r}(a). As the interface lies within the semiconductor shell, there is no Fermi surface mismatch.
 Partial proximitization of a semiconductor wire with a superconductor having a gap $\Delta_s=0.5 \text{ meV}$, results in an induced gap of $\Delta_i=0.15 \text{ meV}$ of the whole system, Figs.\ \ref{3x2r}(b,c).
The wavefunction acquires the angular symmetry of the shell, the localization probability peak however is found to be centralized in the shell, irrespective of the radial asymmetry of proximitization.
 This follows from the boundary condition of no hopping over the inner and outer boundary of the shell, which corresponds to vanishing of the wavefunction in the continuous lattice limit. Note that according to Eq.\ \eqref{zbasis}, the localization probability oscillates along the wire length. The oscillation wavelength is determined by the chemical potential, Fig.\ \ref{3x2r}(d), as the higher energy level increases the frequency. Figs.\ \ref{3x2r}(e,f) show the longitudinal summation of localization probability for the first five positive and negative energy states.  
The boundary conditions force the induced hole component to be equally localized over both the proximitized and non-proximitized parts of the wire shell. In this case, finite size effects\cite{Kiendl_2019,Reeg_2016} dominate over scattering, leading to a uniform electron-hole coherence and an induced gap in the whole shell.
In Fig.\ \ref{deltaar} the induced superconducting gap is shown as a function of the ratio between the non-proximitized and proximitized parts of the shell. The results are found to be independent of the shell thickness, for the given diameter of the wire. The superconducting gap parameter of the proximitized part is set to $\Delta_s= 0.5 \text{ meV}$. The induced superconducting gap of the fully proximitized system is lowered by 10\% due to the applied external magnetic field, $|\vec{B}|=65.8 \text{ mT}$, included in the simulation to lift spin degeneracy. The spin degeneracy is lifted to identify the chemical potential range that includes both spins in the presence of a magnetic field, the effects of which are further studied in Sect.\ \ref{long}.

\FloatBarrier

\section{PARTIAL ANGULAR PROXIMITIZATION}
\label{ang}
\begin{figure}[ht!]
	\centering
	\includegraphics[width=0.48\textwidth]{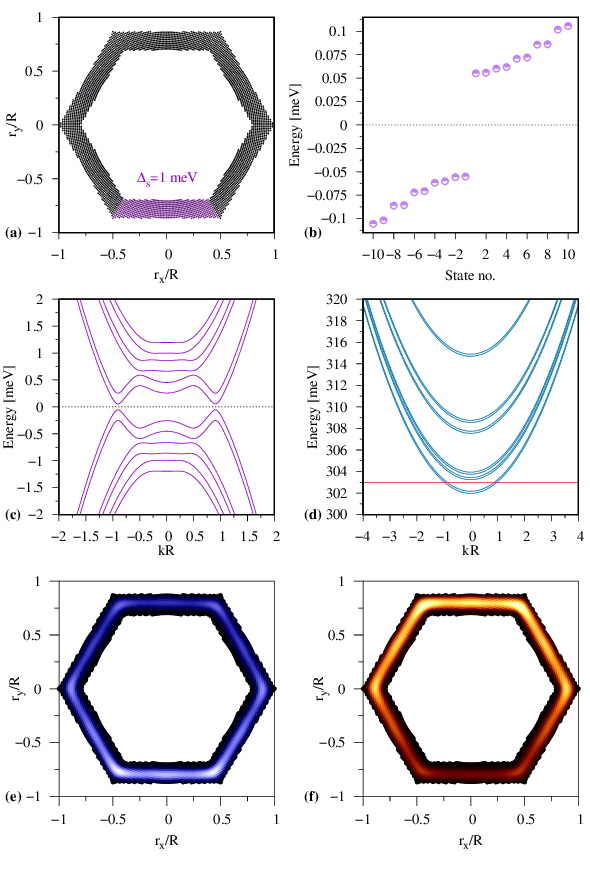}
	\caption{(a) Partial angular proximitization of a hexagonal nanowire shell, single side proximitization with $\Delta_s=1 \text{ meV}$. (b) Finite wire BdG spectrum of the whole system, showing the induced gap $\Delta_i=0.05 \text{ meV}$. (c) Infinite wire BdG spectra. (d) Dispersion and chemical potential (red) of the infinite wire Hamiltonian. (e) Longitudinal summation of localization probability for the hole component $|v|^2$ of the lowest positive energy state, brightness denotes higher localization probability. (f) Corresponding electron component $|u|^2$.}	
	\label{deltaphi}
\end{figure}

Systems where nanowires are proximitized by fabrication of the wire on top of a superconducting slab are common experimental platforms for Majorana physics\cite{Zhang2019,mfcs2014,Stanescu2013}. 
A hexagonal core-shell structure is considered to model such a system where the effective coherence length is smaller than the diameter of the nanowire, such that only a single side can be considered fully proximitized, Fig.\ \ref{deltaphi}(a).
In the fully proximitized part, the gap parameter is set at $\Delta_s=1 \text{ meV}$ and the induced gap obtained is $\Delta_i=0.05 \text{ meV}$, Figs.\ \ref{deltaphi}(b,c). For the first excited positive energy quasiparticle state, in the case of the chemical potential including only the lowest energy band, Fig.\ \ref{deltaphi}(d), the Andreev picture of the proximity effect in uncovered. However, it is only seen by lengthwise summation of localization probability such that the total localization probability is projected onto the wire cross-section, Figs.\ \ref{deltaphi}(e,f).
The fully proximitized part of the semiconductor shell has a hole component localized within it, by definition of the BdG quasiparticle spectrum. Reminiscent of Andreev reflection, the hole components spreads out to the normal conducting part of the shell, Fig.\ \ref{deltaphi}(e). 
The electron localization probability is lowered within the superconducting shell, Fig.\ \ref{deltaphi}(f), in accordance with the view that the superconductor incorporates an electron to form a Cooper pair, and reflects a hole in the process \cite{BTK}. 
The electron-hole coherence results in a near uniformly spread out BdG localization probability over the shell with peaks in the corners, due to corner localization\cite{Sitek15}.
The first negative energy state has the opposite electron-hole localization probability
from Figs.\ \ref{deltaphi}(e,f), as expected from electron-hole symmetry of the system.
Along the length of the wire, the wavefunction localization probability of each state oscillates, Sect.\ \ref{long}, and the symmetry of Figs.\ \ref{deltaphi}(e,f) can be inverted at specific sites. This also happens for the adjacent higher energy states in which the particle-hole coherence is inverted since for a given energy value of the BdG spectra slightly above $\Delta$, there are four states in each band, two electron dominant and the other two hole dominant.

\section{PARTIAL LONGITUDINAL PROXIMITIZATION}
\label{long}
Another possibility of partial proximitization is partial covering of a nanowire longitudinally with a superconductor \cite{Zhang2019,Gul2018}.  A half proximitized wire is considered, such that the superconducting gap is uniform in the whole shell up to half the length of the wire, with $\Delta_S=0.5 \text{ meV}$. 
\begin{figure}[h!]
	\centering
	\includegraphics[width=0.49\textwidth]{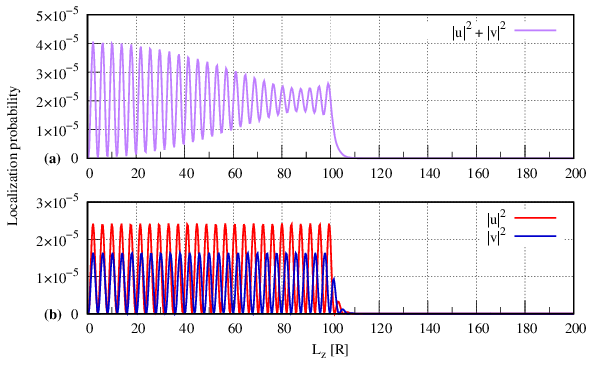}
	\caption{(a) Single corner site localization probability from the composite BdG wavefunction of the lowest energy state of a half proximitized hexagonal wire with no external magnetic field. (b) Corresponding electron and hole components, $|u|^2$ and $|v|^2$ respectively.}
	\label{longiloc}
\end{figure}
Fig.\ \ref{longiloc} shows the electron-hole coherence at a single corner site, for the case of no external magnetic field, of a long hexagonal nanowire with $L=200 \text{ R}$, where the diameter of the wire is 100 nm.  
An exponential decay of the composite BdG localization probability into the non-proximitized part is obtained, Fig.\ \ref{longiloc}(a). This stems from coherence of electron and hole tunneling tails into the non-proximitized half of the wire, Fig.\ \ref{longiloc}(b). Diminishing of the BdG localization probabilty in the proximitized half of the wire is caused by a phase difference between the electron and hole wavefunction components, the $-\pi/2$ phase difference is characteristic of Andreev reflection \cite{Millo_etal,Beenakker91}. 

\begin{figure}[h!]
    \centering
    \includegraphics[width=0.49\textwidth]{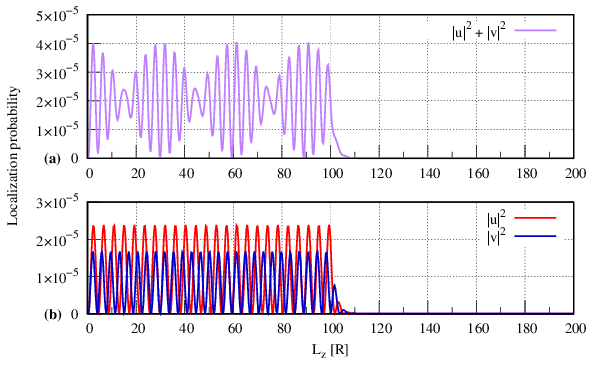}
    \caption{(a) Single corner site localization probability from the composite BdG wavefunction of the lowest energy state of a half proximitized hexagonal wire, for the case of an external magnetic field $|\vec{B}|=65.8 \text{ mT}$. (b) Corresponding electron and hole components, $|u|^2$ and $|v|^2$ respectively.}
    \label{longiloc_B}
\end{figure}

If an external magnetic field is applied, Zeeman splitting gives rise to a difference of the $k$-vectors between the spin-split states, resulting in an additional phase difference between the electron and hole components, Fig.\ \ref{longiloc_B}(b). This phase difference leads to a beating pattern of the BdG localization probability, Fig.\ \ref{longiloc_B}(a).

\begin{figure}[h!]
    \centering
    \includegraphics[width=0.49\textwidth]{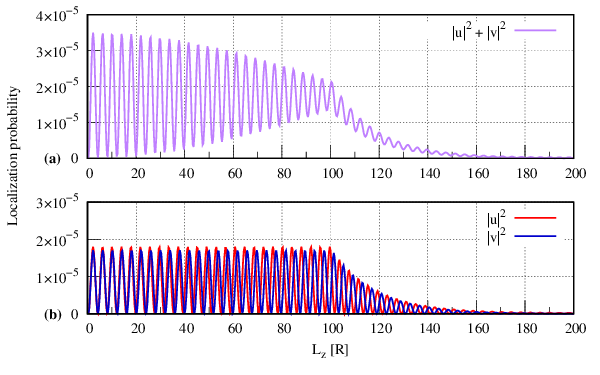}
    \caption{(a) Single corner site localization probability from the composite BdG wavefunction of the lowest energy state of a half proximitized hexagonal wire, $\Delta_S=50 \text{ } \mu\text{eV}$, with no external magnetic field. (b) Corresponding electron and hole components, $|u|^2$ and $|v|^2$ respectively.}
    \label{longiloc_0p1Delta}
\end{figure}

\begin{figure}
    \centering
    \includegraphics[width=0.49\textwidth]{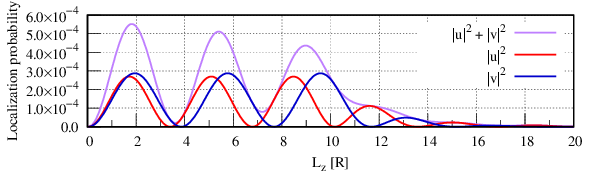}
    \caption{Single corner site localization probability of the composite BdG wavefunction and its corresponding electron $|u|^2$ and hole $|v|^2$ components of the lowest positive energy state.  Wire length is $L=1 \, \mu \text{m}$.}
    \label{longiloc_s}
\end{figure}

\FloatBarrier
In the case of a weaker superconducting gap parameter $\Delta_S=50 \text{ } \mu\text{eV}$, Fig.\ \ref{longiloc_0p1Delta}, the exponential decay into the semiconducting part is enlarged compared with Fig.\ \ref{longiloc}. The gap parameter can thus be seen as an effective potential barrier for the electron-hole coherence.
For a shorter wire with $L= 1 \, \mu \text{m}$ and $\Delta_S=0.5 \text{ meV}$, Fig.\ \ref{longiloc_s}, the exponential decay is less pronounced, compared with Fig.\ \ref{longiloc}. The coherence length is the same but the electron component at the interface is near minimum in phase, rather than at maximum as in the case of the longer wire. 
The wavelength of the wavefunction depends on the Fermi level, and so the length can influence the phase value of the electron and hole components at the interface. For both finite wires, long and short, a direct correspondence with Andreev reflection is found per site of the shell, where the propagation of the quasiparticle wavefunction into the non-proximitized part can be understood in terms of the electron-hole coherence.
\FloatBarrier

\section{DISCUSSION}
The manifestation of Andreev reflection in the electron-hole coherence of the lowest energy quasiparticle states of partially proximitized core-shell nanowires, has been investigated by separately considering radial, angular and longitudinal interfaces of an induced superconducting pairing potential. In the radial case, finite-size effects in the thin shell lead to in-phase electron and hole components, with a symmetric distribution of the total localization probability over the interface. The shell is therefore found to be uniformly gapped, with the superconducting gap of the wire as a whole being suppressed with respect to the proximitized gap. For the angular case, asymmetry of the total localization probability over the interface of the lowest energy state, is shown to correspond to Andreev reflection phenomenology. Including higher energy states results in more complex distributions of the total localization probability, worthy of further investigation, as higher order scattering events come into play.
The most direct correspondence to Andreev reflection is found for the longitudinal interface, where the electron-hole coherence  propagates the quasiparticle wavefunction into the non-proximitized part of the wire.
 This is found for each transverse site, which is effectively a one-dimensional nanowire. The results confirm the applicability of the phenomenology for partially proximitized wires of these dimensions. Andreev reflection as described by BTK\cite{BTK} is a one dimensional picture of electron-hole coherence from scattering at an infinite normal conductor-superconductor (NS) interface, utilizing the Andreev approximation $k_{e/h}^N\approx k_{e/h}^S \approx k_f$, i.e.\ that all wavevectors are evaluated around the Fermi level.  The notion has proven to be very useful for understanding multiple aspects of hybrid superconductor systems in the past decades \cite{Klapwijk2004,AndreevProx} and is still in full use \cite{Ko2022,Hashisaka2021}. It is however an effective single-channel model and limitations of the phenomenology can be expected to come to light with increasing complexity and diversity of fabricated systems \cite{Chalsani_2007,Gupta_2005}.  Significance of normal reflection events for two dimensional interfaces has been shown analytically \cite{Mortensen_1999,Sipr_1997}. The scattering approach has been shown to have its merits in multiple regimes \cite{Beenakker_1997} and even advantages over quasiclassical Green's function approaches \cite{Beenakker_2009}.  However, incorporating self-energy corrections from the superconductor in the Green's function of proximitized nanowires has been shown to capture important renormalization effects on proximity-induced topological superconductivity \cite{TudorSarma_2017}, and has recently been used to quantify the effects of disorder and Fermi surface mismatch in thin-film proximitization \cite{Stanescu_2022}.
The current methodology of numerical diagonalization of the system in a position representation is closer to the scattering approach, providing real-space solutions of the wavefunctions.

\section{CONCLUSIONS}
Low energy physics of the radial, angular and longitudinal superconductor interfaces of proximitized core-shell nanowires has been explored using the Bogoliubov-de Gennes equations. Partial proximitization is considered to quantify the effects of a reduced coherence length and to investigate the Andreev reflection interpretation of proximity induced superconductivity. For a thin shell, boundary conditions are found to impose symmetry of the localization probability in spite of partial radial proximitization. In the case of a hexagonal wire with a single proximitized side, it is only by lengthwise summation of localization probability that the essence of Andreev reflection can be seen. For a longitudinally half proximitized wire, electron-hole coherence is explicitly shown to propagate the quasiparticle wavefunction into the non-proximitized part of the wire. Application of an external magnetic field directed along the wire axis is found to cause amplitude modulation of the quasiparticle wavefunction.
Correspondence with Andreev reflection is obtained per site of the shell, from considering the localization probability as a function of length in the core-shell nanowire. The results show in what way Andreev reflection is compatible with electron-hole coherence at the various interfaces that can arise in proximitized core-shell nanowire systems.

\begin{acknowledgments}
This research was supported by the Reykjavik University Research Fund, Project No.\ 218043 and the Icelandic Research Fund, Grant No. 206568-051. We thank Thomas Schäpers and Vidar Gudmundsson for discussions. 
\end{acknowledgments}
\bibliographystyle{apsrev4-1}
\bibliography{CSNreferences_v08_doirm}

\begin{thebibliography}{118}%
\makeatletter
\providecommand \@ifxundefined [1]{%
 \@ifx{#1\undefined}
}%
\providecommand \@ifnum [1]{%
 \ifnum #1\expandafter \@firstoftwo
 \else \expandafter \@secondoftwo
 \fi
}%
\providecommand \@ifx [1]{%
 \ifx #1\expandafter \@firstoftwo
 \else \expandafter \@secondoftwo
 \fi
}%
\providecommand \natexlab [1]{#1}%
\providecommand \enquote  [1]{``#1''}%
\providecommand \bibnamefont  [1]{#1}%
\providecommand \bibfnamefont [1]{#1}%
\providecommand \citenamefont [1]{#1}%
\providecommand \href@noop [0]{\@secondoftwo}%
\providecommand \href [0]{\begingroup \@sanitize@url \@href}%
\providecommand \@href[1]{\@@startlink{#1}\@@href}%
\providecommand \@@href[1]{\endgroup#1\@@endlink}%
\providecommand \@sanitize@url [0]{\catcode `\\12\catcode `\$12\catcode
  `\&12\catcode `\#12\catcode `\^12\catcode `\_12\catcode `\%12\relax}%
\providecommand \@@startlink[1]{}%
\providecommand \@@endlink[0]{}%
\providecommand \url  [0]{\begingroup\@sanitize@url \@url }%
\providecommand \@url [1]{\endgroup\@href {#1}{\urlprefix }}%
\providecommand \urlprefix  [0]{URL }%
\providecommand \Eprint [0]{\href }%
\providecommand \doibase [0]{http://dx.doi.org/}%
\providecommand \selectlanguage [0]{\@gobble}%
\providecommand \bibinfo  [0]{\@secondoftwo}%
\providecommand \bibfield  [0]{\@secondoftwo}%
\providecommand \translation [1]{[#1]}%
\providecommand \BibitemOpen [0]{}%
\providecommand \bibitemStop [0]{}%
\providecommand \bibitemNoStop [0]{.\EOS\space}%
\providecommand \EOS [0]{\spacefactor3000\relax}%
\providecommand \BibitemShut  [1]{\csname bibitem#1\endcsname}%
\let\auto@bib@innerbib\@empty
\bibitem [{\citenamefont {Hays}\ \emph {et~al.}(2021)\citenamefont {Hays},
  \citenamefont {Fatemi}, \citenamefont {Bouman}, \citenamefont {Cerrillo},
  \citenamefont {Diamond}, \citenamefont {Serniak}, \citenamefont {Connolly},
  \citenamefont {Krogstrup}, \citenamefont {Nygård}, \citenamefont {Yeyati},
  \citenamefont {Geresdi},\ and\ \citenamefont {Devoret}}]{Hays_21}%
  \BibitemOpen
  \bibfield  {author} {\bibinfo {author} {\bibfnamefont {M.}~\bibnamefont
  {Hays}}, \bibinfo {author} {\bibfnamefont {V.}~\bibnamefont {Fatemi}},
  \bibinfo {author} {\bibfnamefont {D.}~\bibnamefont {Bouman}}, \bibinfo
  {author} {\bibfnamefont {J.}~\bibnamefont {Cerrillo}}, \bibinfo {author}
  {\bibfnamefont {S.}~\bibnamefont {Diamond}}, \bibinfo {author} {\bibfnamefont
  {K.}~\bibnamefont {Serniak}}, \bibinfo {author} {\bibfnamefont
  {T.}~\bibnamefont {Connolly}}, \bibinfo {author} {\bibfnamefont
  {P.}~\bibnamefont {Krogstrup}}, \bibinfo {author} {\bibfnamefont
  {J.}~\bibnamefont {Nygård}}, \bibinfo {author} {\bibfnamefont {A.~L.}\
  \bibnamefont {Yeyati}}, \bibinfo {author} {\bibfnamefont {A.}~\bibnamefont
  {Geresdi}}, \ and\ \bibinfo {author} {\bibfnamefont {M.~H.}\ \bibnamefont
  {Devoret}},\ }\href {\doibase 10.1126/science.abf0345} {\bibfield  {journal}
  {\bibinfo  {journal} {Science}\ }\textbf {\bibinfo {volume} {373}},\ \bibinfo
  {pages} {430} (\bibinfo {year} {2021})},\ \Eprint
  {http://arxiv.org/abs/https://www.science.org/doi/pdf/10.1126/science.abf0345}
  {https://www.science.org/doi/pdf/10.1126/science.abf0345} \BibitemShut
  {NoStop}%
\bibitem [{\citenamefont {Aguado}(2020)}]{Aguado2020}%
  \BibitemOpen
  \bibfield  {author} {\bibinfo {author} {\bibfnamefont {R.}~\bibnamefont
  {Aguado}},\ }\href {\doibase 10.1063/5.0024124} {\bibfield  {journal}
  {\bibinfo  {journal} {Applied Physics Letters}\ }\textbf {\bibinfo {volume}
  {117}},\ \bibinfo {pages} {240501} (\bibinfo {year} {2020})},\ \Eprint
  {http://arxiv.org/abs/https://doi.org/10.1063/5.0024124}
  {https://doi.org/10.1063/5.0024124} \BibitemShut {NoStop}%
\bibitem [{\citenamefont {Benito}\ and\ \citenamefont
  {Burkard}(2020)}]{Benito2020}%
  \BibitemOpen
  \bibfield  {author} {\bibinfo {author} {\bibfnamefont {M.}~\bibnamefont
  {Benito}}\ and\ \bibinfo {author} {\bibfnamefont {G.}~\bibnamefont
  {Burkard}},\ }\href@noop {} {\bibfield  {journal} {\bibinfo  {journal}
  {Applied Physics Letters}\ }\textbf {\bibinfo {volume} {116}},\ \bibinfo
  {pages} {190502} (\bibinfo {year} {2020})}\BibitemShut {NoStop}%
\bibitem [{\citenamefont {Andreev}(1964)}]{Andreev}%
  \BibitemOpen
  \bibfield  {author} {\bibinfo {author} {\bibfnamefont {A.}~\bibnamefont
  {Andreev}},\ }\href@noop {} {\bibfield  {journal} {\bibinfo  {journal}
  {Journal of Experimental and Theoretical Physics}\ }\textbf {\bibinfo
  {volume} {46}},\ \bibinfo {pages} {1823} (\bibinfo {year}
  {1964})}\BibitemShut {NoStop}%
\bibitem [{\citenamefont {Pannetier}\ and\ \citenamefont
  {Courtois}(2000)}]{AndreevProx}%
  \BibitemOpen
  \bibfield  {author} {\bibinfo {author} {\bibfnamefont {B.}~\bibnamefont
  {Pannetier}}\ and\ \bibinfo {author} {\bibfnamefont {H.}~\bibnamefont
  {Courtois}},\ }\href@noop {} {\bibfield  {journal} {\bibinfo  {journal}
  {Journal of Low Temperature Physics}\ }\textbf {\bibinfo {volume} {118}},\
  \bibinfo {pages} {599} (\bibinfo {year} {2000})}\BibitemShut {NoStop}%
\bibitem [{\citenamefont {Blonder}\ \emph {et~al.}(1982)\citenamefont
  {Blonder}, \citenamefont {Tinkham},\ and\ \citenamefont {Klapwijk}}]{BTK}%
  \BibitemOpen
  \bibfield  {author} {\bibinfo {author} {\bibfnamefont {G.~E.}\ \bibnamefont
  {Blonder}}, \bibinfo {author} {\bibfnamefont {M.}~\bibnamefont {Tinkham}}, \
  and\ \bibinfo {author} {\bibfnamefont {T.~M.}\ \bibnamefont {Klapwijk}},\
  }\href@noop {} {\bibfield  {journal} {\bibinfo  {journal} {Phys. Rev. B}\
  }\textbf {\bibinfo {volume} {25}},\ \bibinfo {pages} {4515} (\bibinfo {year}
  {1982})}\BibitemShut {NoStop}%
\bibitem [{\citenamefont {Setiawan}\ \emph {et~al.}(2019)\citenamefont
  {Setiawan}, \citenamefont {Wu},\ and\ \citenamefont {Levin}}]{TSC_prox}%
  \BibitemOpen
  \bibfield  {author} {\bibinfo {author} {\bibfnamefont {F.}~\bibnamefont
  {Setiawan}}, \bibinfo {author} {\bibfnamefont {C.-T.}\ \bibnamefont {Wu}}, \
  and\ \bibinfo {author} {\bibfnamefont {K.}~\bibnamefont {Levin}},\ }\href
  {\doibase 10.1103/PhysRevB.99.174511} {\bibfield  {journal} {\bibinfo
  {journal} {Phys. Rev. B}\ }\textbf {\bibinfo {volume} {99}},\ \bibinfo
  {pages} {174511} (\bibinfo {year} {2019})}\BibitemShut {NoStop}%
\bibitem [{\citenamefont {Stanescu}\ \emph {et~al.}(2014)\citenamefont
  {Stanescu}, \citenamefont {Lutchyn},\ and\ \citenamefont {{Das
  Sarma}}}]{Tudor2014}%
  \BibitemOpen
  \bibfield  {author} {\bibinfo {author} {\bibfnamefont {T.~D.}\ \bibnamefont
  {Stanescu}}, \bibinfo {author} {\bibfnamefont {R.~M.}\ \bibnamefont
  {Lutchyn}}, \ and\ \bibinfo {author} {\bibfnamefont {S.}~\bibnamefont {{Das
  Sarma}}},\ }\href {\doibase 10.1103/PhysRevB.90.085302} {\bibfield  {journal}
  {\bibinfo  {journal} {Phys. Rev. B}\ }\textbf {\bibinfo {volume} {90}},\
  \bibinfo {pages} {085302} (\bibinfo {year} {2014})}\BibitemShut {NoStop}%
\bibitem [{\citenamefont {Mu\~noz}\ \emph {et~al.}(2013)\citenamefont
  {Mu\~noz}, \citenamefont {Covaci},\ and\ \citenamefont
  {Peeters}}]{Graphene_2013}%
  \BibitemOpen
  \bibfield  {author} {\bibinfo {author} {\bibfnamefont {W.~A.}\ \bibnamefont
  {Mu\~noz}}, \bibinfo {author} {\bibfnamefont {L.}~\bibnamefont {Covaci}}, \
  and\ \bibinfo {author} {\bibfnamefont {F.~M.}\ \bibnamefont {Peeters}},\
  }\href {\doibase 10.1103/PhysRevB.87.134509} {\bibfield  {journal} {\bibinfo
  {journal} {Phys. Rev. B}\ }\textbf {\bibinfo {volume} {87}},\ \bibinfo
  {pages} {134509} (\bibinfo {year} {2013})}\BibitemShut {NoStop}%
\bibitem [{\citenamefont {Zha}\ \emph {et~al.}(2010)\citenamefont {Zha},
  \citenamefont {Covaci}, \citenamefont {Zhou},\ and\ \citenamefont
  {Peeters}}]{Zha_2010}%
  \BibitemOpen
  \bibfield  {author} {\bibinfo {author} {\bibfnamefont {G.-Q.}\ \bibnamefont
  {Zha}}, \bibinfo {author} {\bibfnamefont {L.}~\bibnamefont {Covaci}},
  \bibinfo {author} {\bibfnamefont {S.-P.}\ \bibnamefont {Zhou}}, \ and\
  \bibinfo {author} {\bibfnamefont {F.~M.}\ \bibnamefont {Peeters}},\ }\href
  {\doibase 10.1103/PhysRevB.82.140502} {\bibfield  {journal} {\bibinfo
  {journal} {Phys. Rev. B}\ }\textbf {\bibinfo {volume} {82}},\ \bibinfo
  {pages} {140502(R)} (\bibinfo {year} {2010})}\BibitemShut {NoStop}%
\bibitem [{\citenamefont {Klapwijk}(2004)}]{Klapwijk2004}%
  \BibitemOpen
  \bibfield  {author} {\bibinfo {author} {\bibfnamefont {T.~M.}\ \bibnamefont
  {Klapwijk}},\ }\href {\doibase 10.1007/s10948-004-0773-0} {\bibfield
  {journal} {\bibinfo  {journal} {Journal of Superconductivity}\ }\textbf
  {\bibinfo {volume} {17}},\ \bibinfo {pages} {593} (\bibinfo {year}
  {2004})}\BibitemShut {NoStop}%
\bibitem [{\citenamefont {Jacobs}\ \emph {et~al.}(1999)\citenamefont {Jacobs},
  \citenamefont {K{\"u}mmel},\ and\ \citenamefont {Plehn}}]{JACOBS1999669}%
  \BibitemOpen
  \bibfield  {author} {\bibinfo {author} {\bibfnamefont {A.}~\bibnamefont
  {Jacobs}}, \bibinfo {author} {\bibfnamefont {R.}~\bibnamefont {K{\"u}mmel}},
  \ and\ \bibinfo {author} {\bibfnamefont {H.}~\bibnamefont {Plehn}},\ }\href
  {\doibase 10.1006/spmi.1999.0718} {\bibfield  {journal} {\bibinfo  {journal}
  {Superlattices and Microstructures}\ }\textbf {\bibinfo {volume} {25}},\
  \bibinfo {pages} {669} (\bibinfo {year} {1999})}\BibitemShut {NoStop}%
\bibitem [{\citenamefont {D'yachenko}\ and\ \citenamefont
  {Kochergin}(1991)}]{osti_5167845}%
  \BibitemOpen
  \bibfield  {author} {\bibinfo {author} {\bibfnamefont {A.~I.}\ \bibnamefont
  {D'yachenko}}\ and\ \bibinfo {author} {\bibfnamefont {I.~V.}\ \bibnamefont
  {Kochergin}},\ }\href {\doibase 10.1007/BF00683608} {\bibfield  {journal}
  {\bibinfo  {journal} {J. Low Temp. Phys.}\ }\textbf {\bibinfo {volume}
  {84:3-4}} (\bibinfo {year} {1991}),\ 10.1007/BF00683608}\BibitemShut
  {NoStop}%
\bibitem [{\citenamefont {van Haesendonck}\ \emph {et~al.}(1981)\citenamefont
  {van Haesendonck}, \citenamefont {den Dries}, \citenamefont {Bruynseraede},\
  and\ \citenamefont {Gilabert}}]{Haesendonck_1981}%
  \BibitemOpen
  \bibfield  {author} {\bibinfo {author} {\bibfnamefont {C.}~\bibnamefont {van
  Haesendonck}}, \bibinfo {author} {\bibfnamefont {L.~V.}\ \bibnamefont {den
  Dries}}, \bibinfo {author} {\bibfnamefont {Y.}~\bibnamefont {Bruynseraede}},
  \ and\ \bibinfo {author} {\bibfnamefont {A.}~\bibnamefont {Gilabert}},\
  }\href@noop {} {\bibfield  {journal} {\bibinfo  {journal} {Journal of Physics
  F: Metal Physics}\ }\textbf {\bibinfo {volume} {11}},\ \bibinfo {pages}
  {2381} (\bibinfo {year} {1981})}\BibitemShut {NoStop}%
\bibitem [{\citenamefont {Meissner}(1971)}]{Meissner71}%
  \BibitemOpen
  \bibfield  {author} {\bibinfo {author} {\bibfnamefont {H.}~\bibnamefont
  {Meissner}},\ }\href {https://ntrs.nasa.gov/citations/19720004977} {\bibfield
   {journal} {\bibinfo  {journal} {Stevens Institute of Technology}\ }
  (\bibinfo {year} {1971})}\BibitemShut {NoStop}%
\bibitem [{\citenamefont {Clarke}(1968)}]{clarke:jpa-00213516}%
  \BibitemOpen
  \bibfield  {author} {\bibinfo {author} {\bibfnamefont {J.}~\bibnamefont
  {Clarke}},\ }\href {\doibase 10.1051/jphyscol:1968201} {\bibfield  {journal}
  {\bibinfo  {journal} {{Journal de Physique Colloques}}\ }\textbf {\bibinfo
  {volume} {29}},\ \bibinfo {pages} {C2} (\bibinfo {year} {1968})}\BibitemShut
  {NoStop}%
\bibitem [{\citenamefont {Prada}\ \emph {et~al.}(2020)\citenamefont {Prada},
  \citenamefont {San-Jose}, \citenamefont {de~Moor}, \citenamefont {Geresdi},
  \citenamefont {Lee}, \citenamefont {Klinovaja}, \citenamefont {Loss},
  \citenamefont {Nyg{\aa}rd}, \citenamefont {Aguado},\ and\ \citenamefont
  {Kouwenhoven}}]{Elsa2020}%
  \BibitemOpen
  \bibfield  {author} {\bibinfo {author} {\bibfnamefont {E.}~\bibnamefont
  {Prada}}, \bibinfo {author} {\bibfnamefont {P.}~\bibnamefont {San-Jose}},
  \bibinfo {author} {\bibfnamefont {M.~W.~A.}\ \bibnamefont {de~Moor}},
  \bibinfo {author} {\bibfnamefont {A.}~\bibnamefont {Geresdi}}, \bibinfo
  {author} {\bibfnamefont {E.~J.~H.}\ \bibnamefont {Lee}}, \bibinfo {author}
  {\bibfnamefont {J.}~\bibnamefont {Klinovaja}}, \bibinfo {author}
  {\bibfnamefont {D.}~\bibnamefont {Loss}}, \bibinfo {author} {\bibfnamefont
  {J.}~\bibnamefont {Nyg{\aa}rd}}, \bibinfo {author} {\bibfnamefont
  {R.}~\bibnamefont {Aguado}}, \ and\ \bibinfo {author} {\bibfnamefont {L.~P.}\
  \bibnamefont {Kouwenhoven}},\ }\href {\doibase 10.1038/s42254-020-0228-y}
  {\bibfield  {journal} {\bibinfo  {journal} {Nature Reviews Physics}\ }\textbf
  {\bibinfo {volume} {2}},\ \bibinfo {pages} {575} (\bibinfo {year}
  {2020})}\BibitemShut {NoStop}%
\bibitem [{\citenamefont {J{\"u}nger}\ \emph {et~al.}(2019)\citenamefont
  {J{\"u}nger}, \citenamefont {Baumgartner}, \citenamefont {Delagrange},
  \citenamefont {Chevallier}, \citenamefont {Lehmann}, \citenamefont {Nilsson},
  \citenamefont {Dick}, \citenamefont {Thelander},\ and\ \citenamefont
  {Sch{\"o}nenberger}}]{prox2019_junger}%
  \BibitemOpen
  \bibfield  {author} {\bibinfo {author} {\bibfnamefont {C.}~\bibnamefont
  {J{\"u}nger}}, \bibinfo {author} {\bibfnamefont {A.}~\bibnamefont
  {Baumgartner}}, \bibinfo {author} {\bibfnamefont {R.}~\bibnamefont
  {Delagrange}}, \bibinfo {author} {\bibfnamefont {D.}~\bibnamefont
  {Chevallier}}, \bibinfo {author} {\bibfnamefont {S.}~\bibnamefont {Lehmann}},
  \bibinfo {author} {\bibfnamefont {M.}~\bibnamefont {Nilsson}}, \bibinfo
  {author} {\bibfnamefont {K.~A.}\ \bibnamefont {Dick}}, \bibinfo {author}
  {\bibfnamefont {C.}~\bibnamefont {Thelander}}, \ and\ \bibinfo {author}
  {\bibfnamefont {C.}~\bibnamefont {Sch{\"o}nenberger}},\ }\href {\doibase
  10.1038/s42005-019-0162-4} {\bibfield  {journal} {\bibinfo  {journal}
  {Communications Physics}\ }\textbf {\bibinfo {volume} {2}},\ \bibinfo {pages}
  {76} (\bibinfo {year} {2019})}\BibitemShut {NoStop}%
\bibitem [{\citenamefont {Aguado}(2017)}]{Review_Agauado}%
  \BibitemOpen
  \bibfield  {author} {\bibinfo {author} {\bibfnamefont {R.}~\bibnamefont
  {Aguado}},\ }\href {\doibase 10.1393/ncr/i2017-10141-9} {\bibfield  {journal}
  {\bibinfo  {journal} {La Rivista del Nuovo Cimento}\ ,\ \bibinfo {pages}
  {523}} (\bibinfo {year} {2017})}\BibitemShut {NoStop}%
\bibitem [{\citenamefont {Stanescu}(2017)}]{TudorBok}%
  \BibitemOpen
  \bibfield  {author} {\bibinfo {author} {\bibfnamefont {T.~D.}\ \bibnamefont
  {Stanescu}},\ }\href@noop {} {\emph {\bibinfo {title} {Introduction to
  Topological Quantum Matter \& Quantum Computation}}}\ (\bibinfo  {publisher}
  {CRC Press: London.},\ \bibinfo {year} {2017})\BibitemShut {NoStop}%
\bibitem [{\citenamefont {Alicea}(2010)}]{Alicea_2010}%
  \BibitemOpen
  \bibfield  {author} {\bibinfo {author} {\bibfnamefont {J.}~\bibnamefont
  {Alicea}},\ }\href {\doibase 10.1103/PhysRevB.81.125318} {\bibfield
  {journal} {\bibinfo  {journal} {Phys. Rev. B}\ }\textbf {\bibinfo {volume}
  {81}},\ \bibinfo {pages} {125318} (\bibinfo {year} {2010})}\BibitemShut
  {NoStop}%
\bibitem [{\citenamefont {Sau}\ \emph {et~al.}(2010)\citenamefont {Sau},
  \citenamefont {Lutchyn}, \citenamefont {Tewari},\ and\ \citenamefont
  {Das~Sarma}}]{Sau2010}%
  \BibitemOpen
  \bibfield  {author} {\bibinfo {author} {\bibfnamefont {J.~D.}\ \bibnamefont
  {Sau}}, \bibinfo {author} {\bibfnamefont {R.~M.}\ \bibnamefont {Lutchyn}},
  \bibinfo {author} {\bibfnamefont {S.}~\bibnamefont {Tewari}}, \ and\ \bibinfo
  {author} {\bibfnamefont {S.}~\bibnamefont {Das~Sarma}},\ }\href {\doibase
  10.1103/PhysRevLett.104.040502} {\bibfield  {journal} {\bibinfo  {journal}
  {Phys. Rev. Lett.}\ }\textbf {\bibinfo {volume} {104}},\ \bibinfo {pages}
  {040502} (\bibinfo {year} {2010})}\BibitemShut {NoStop}%
\bibitem [{\citenamefont {Bl{\"o}mers}\ \emph {et~al.}(2013)\citenamefont
  {Bl{\"o}mers}, \citenamefont {Rieger}, \citenamefont {Zellekens},
  \citenamefont {Haas}, \citenamefont {Lepsa}, \citenamefont {Hardtdegen},
  \citenamefont {G{\"u}l}, \citenamefont {Demarina}, \citenamefont
  {Gr{\"u}tzmacher}, \citenamefont {L{\"u}th},\ and\ \citenamefont
  {Sch{\"a}pers}}]{Blomers13}%
  \BibitemOpen
  \bibfield  {author} {\bibinfo {author} {\bibfnamefont {C.}~\bibnamefont
  {Bl{\"o}mers}}, \bibinfo {author} {\bibfnamefont {T.}~\bibnamefont {Rieger}},
  \bibinfo {author} {\bibfnamefont {P.}~\bibnamefont {Zellekens}}, \bibinfo
  {author} {\bibfnamefont {F.}~\bibnamefont {Haas}}, \bibinfo {author}
  {\bibfnamefont {M.~I.}\ \bibnamefont {Lepsa}}, \bibinfo {author}
  {\bibfnamefont {H.}~\bibnamefont {Hardtdegen}}, \bibinfo {author}
  {\bibfnamefont {{\"O}.}~\bibnamefont {G{\"u}l}}, \bibinfo {author}
  {\bibfnamefont {N.}~\bibnamefont {Demarina}}, \bibinfo {author}
  {\bibfnamefont {D.}~\bibnamefont {Gr{\"u}tzmacher}}, \bibinfo {author}
  {\bibfnamefont {H.}~\bibnamefont {L{\"u}th}}, \ and\ \bibinfo {author}
  {\bibfnamefont {T.}~\bibnamefont {Sch{\"a}pers}},\ }\href@noop {} {\bibfield
  {journal} {\bibinfo  {journal} {Nanotechnology}\ }\textbf {\bibinfo {volume}
  {24}},\ \bibinfo {pages} {035203} (\bibinfo {year} {2013})}\BibitemShut
  {NoStop}%
\bibitem [{\citenamefont {Rieger}\ \emph {et~al.}(2012)\citenamefont {Rieger},
  \citenamefont {Luysberg}, \citenamefont {Sch{\"a}pers}, \citenamefont
  {Gr{\"u}tzmacher},\ and\ \citenamefont {Lepsa}}]{Rieger12}%
  \BibitemOpen
  \bibfield  {author} {\bibinfo {author} {\bibfnamefont {T.}~\bibnamefont
  {Rieger}}, \bibinfo {author} {\bibfnamefont {M.}~\bibnamefont {Luysberg}},
  \bibinfo {author} {\bibfnamefont {T.}~\bibnamefont {Sch{\"a}pers}}, \bibinfo
  {author} {\bibfnamefont {D.}~\bibnamefont {Gr{\"u}tzmacher}}, \ and\ \bibinfo
  {author} {\bibfnamefont {M.~I.}\ \bibnamefont {Lepsa}},\ }\href@noop {}
  {\bibfield  {journal} {\bibinfo  {journal} {Nano Letters}\ }\textbf {\bibinfo
  {volume} {12}},\ \bibinfo {pages} {5559} (\bibinfo {year}
  {2012})}\BibitemShut {NoStop}%
\bibitem [{\citenamefont {Haas}\ \emph {et~al.}(2013)\citenamefont {Haas},
  \citenamefont {Sladek}, \citenamefont {Winden}, \citenamefont {von~der Ahe},
  \citenamefont {Weirich}, \citenamefont {Rieger}, \citenamefont {L{\"u}th},
  \citenamefont {Gr{\"u}tzmacher}, \citenamefont {Sch{\"a}pers},\ and\
  \citenamefont {Hardtdegen}}]{Haas13}%
  \BibitemOpen
  \bibfield  {author} {\bibinfo {author} {\bibfnamefont {F.}~\bibnamefont
  {Haas}}, \bibinfo {author} {\bibfnamefont {K.}~\bibnamefont {Sladek}},
  \bibinfo {author} {\bibfnamefont {A.}~\bibnamefont {Winden}}, \bibinfo
  {author} {\bibfnamefont {M.}~\bibnamefont {von~der Ahe}}, \bibinfo {author}
  {\bibfnamefont {T.~E.}\ \bibnamefont {Weirich}}, \bibinfo {author}
  {\bibfnamefont {T.}~\bibnamefont {Rieger}}, \bibinfo {author} {\bibfnamefont
  {H.}~\bibnamefont {L{\"u}th}}, \bibinfo {author} {\bibfnamefont
  {D.}~\bibnamefont {Gr{\"u}tzmacher}}, \bibinfo {author} {\bibfnamefont
  {T.}~\bibnamefont {Sch{\"a}pers}}, \ and\ \bibinfo {author} {\bibfnamefont
  {H.}~\bibnamefont {Hardtdegen}},\ }\href@noop {} {\bibfield  {journal}
  {\bibinfo  {journal} {Nanotechnology}\ }\textbf {\bibinfo {volume} {24}},\
  \bibinfo {pages} {085603} (\bibinfo {year} {2013})}\BibitemShut {NoStop}%
\bibitem [{\citenamefont {Funk}\ \emph {et~al.}(2013)\citenamefont {Funk},
  \citenamefont {Royo}, \citenamefont {Zardo}, \citenamefont {Rudolph},
  \citenamefont {Morkötter}, \citenamefont {Mayer}, \citenamefont {Becker},
  \citenamefont {Bechtold}, \citenamefont {Matich}, \citenamefont {Döblinger},
  \citenamefont {Bichler}, \citenamefont {Koblmüller}, \citenamefont {Finley},
  \citenamefont {Bertoni}, \citenamefont {Goldoni},\ and\ \citenamefont
  {Abstreiter}}]{Funk13}%
  \BibitemOpen
  \bibfield  {author} {\bibinfo {author} {\bibfnamefont {S.}~\bibnamefont
  {Funk}}, \bibinfo {author} {\bibfnamefont {M.}~\bibnamefont {Royo}}, \bibinfo
  {author} {\bibfnamefont {I.}~\bibnamefont {Zardo}}, \bibinfo {author}
  {\bibfnamefont {D.}~\bibnamefont {Rudolph}}, \bibinfo {author} {\bibfnamefont
  {S.}~\bibnamefont {Morkötter}}, \bibinfo {author} {\bibfnamefont
  {B.}~\bibnamefont {Mayer}}, \bibinfo {author} {\bibfnamefont
  {J.}~\bibnamefont {Becker}}, \bibinfo {author} {\bibfnamefont
  {A.}~\bibnamefont {Bechtold}}, \bibinfo {author} {\bibfnamefont
  {S.}~\bibnamefont {Matich}}, \bibinfo {author} {\bibfnamefont
  {M.}~\bibnamefont {Döblinger}}, \bibinfo {author} {\bibfnamefont
  {M.}~\bibnamefont {Bichler}}, \bibinfo {author} {\bibfnamefont
  {G.}~\bibnamefont {Koblmüller}}, \bibinfo {author} {\bibfnamefont {J.~J.}\
  \bibnamefont {Finley}}, \bibinfo {author} {\bibfnamefont {A.}~\bibnamefont
  {Bertoni}}, \bibinfo {author} {\bibfnamefont {G.}~\bibnamefont {Goldoni}}, \
  and\ \bibinfo {author} {\bibfnamefont {G.}~\bibnamefont {Abstreiter}},\
  }\href@noop {} {\bibfield  {journal} {\bibinfo  {journal} {Nano Letters}\
  }\textbf {\bibinfo {volume} {13}},\ \bibinfo {pages} {6189} (\bibinfo {year}
  {2013})}\BibitemShut {NoStop}%
\bibitem [{\citenamefont {Erhard}\ \emph {et~al.}(2015)\citenamefont {Erhard},
  \citenamefont {Zenger}, \citenamefont {Morkötter}, \citenamefont {Rudolph},
  \citenamefont {Weiss}, \citenamefont {Krenner}, \citenamefont {Karl},
  \citenamefont {Abstreiter}, \citenamefont {Finley}, \citenamefont
  {Koblm{\"u}ller},\ and\ \citenamefont {Holleitner}}]{Erhard15}%
  \BibitemOpen
  \bibfield  {author} {\bibinfo {author} {\bibfnamefont {N.}~\bibnamefont
  {Erhard}}, \bibinfo {author} {\bibfnamefont {S.}~\bibnamefont {Zenger}},
  \bibinfo {author} {\bibfnamefont {S.}~\bibnamefont {Morkötter}}, \bibinfo
  {author} {\bibfnamefont {D.}~\bibnamefont {Rudolph}}, \bibinfo {author}
  {\bibfnamefont {M.}~\bibnamefont {Weiss}}, \bibinfo {author} {\bibfnamefont
  {H.~J.}\ \bibnamefont {Krenner}}, \bibinfo {author} {\bibfnamefont
  {H.}~\bibnamefont {Karl}}, \bibinfo {author} {\bibfnamefont {G.}~\bibnamefont
  {Abstreiter}}, \bibinfo {author} {\bibfnamefont {J.~J.}\ \bibnamefont
  {Finley}}, \bibinfo {author} {\bibfnamefont {G.}~\bibnamefont
  {Koblm{\"u}ller}}, \ and\ \bibinfo {author} {\bibfnamefont {A.~W.}\
  \bibnamefont {Holleitner}},\ }\href@noop {} {\bibfield  {journal} {\bibinfo
  {journal} {Nano Letters}\ }\textbf {\bibinfo {volume} {15}},\ \bibinfo
  {pages} {6869} (\bibinfo {year} {2015})}\BibitemShut {NoStop}%
\bibitem [{\citenamefont {Wei{\ss}}\ \emph {et~al.}(2014)\citenamefont
  {Wei{\ss}}, \citenamefont {Kinzel}, \citenamefont {Sch{\"u}lein},
  \citenamefont {Heigl}, \citenamefont {Rudolph}, \citenamefont
  {Mork{\"o}tter}, \citenamefont {D{\"o}blinger}, \citenamefont {Bichler},
  \citenamefont {Abstreiter}, \citenamefont {Finley}, \citenamefont
  {Koblm{\"u}ller}, \citenamefont {Wixforth},\ and\ \citenamefont
  {Krenner}}]{Weiss14b}%
  \BibitemOpen
  \bibfield  {author} {\bibinfo {author} {\bibfnamefont {M.}~\bibnamefont
  {Wei{\ss}}}, \bibinfo {author} {\bibfnamefont {J.~B.}\ \bibnamefont
  {Kinzel}}, \bibinfo {author} {\bibfnamefont {F.~J.~R.}\ \bibnamefont
  {Sch{\"u}lein}}, \bibinfo {author} {\bibfnamefont {M.}~\bibnamefont {Heigl}},
  \bibinfo {author} {\bibfnamefont {D.}~\bibnamefont {Rudolph}}, \bibinfo
  {author} {\bibfnamefont {S.}~\bibnamefont {Mork{\"o}tter}}, \bibinfo {author}
  {\bibfnamefont {M.}~\bibnamefont {D{\"o}blinger}}, \bibinfo {author}
  {\bibfnamefont {M.}~\bibnamefont {Bichler}}, \bibinfo {author} {\bibfnamefont
  {G.}~\bibnamefont {Abstreiter}}, \bibinfo {author} {\bibfnamefont {J.~J.}\
  \bibnamefont {Finley}}, \bibinfo {author} {\bibfnamefont {G.}~\bibnamefont
  {Koblm{\"u}ller}}, \bibinfo {author} {\bibfnamefont {A.}~\bibnamefont
  {Wixforth}}, \ and\ \bibinfo {author} {\bibfnamefont {H.~J.}\ \bibnamefont
  {Krenner}},\ }\href@noop {} {\bibfield  {journal} {\bibinfo  {journal} {Nano
  Letters}\ }\textbf {\bibinfo {volume} {14}},\ \bibinfo {pages} {2256}
  (\bibinfo {year} {2014})}\BibitemShut {NoStop}%
\bibitem [{\citenamefont {Jadczak}\ \emph {et~al.}(2014)\citenamefont
  {Jadczak}, \citenamefont {Plochocka}, \citenamefont {Mitioglu}, \citenamefont
  {Breslavetz}, \citenamefont {Royo}, \citenamefont {Bertoni}, \citenamefont
  {Goldoni}, \citenamefont {Smolenski}, \citenamefont {Kossacki}, \citenamefont
  {Kretinin}, \citenamefont {Shtrikman},\ and\ \citenamefont
  {Maude}}]{Jadczak14}%
  \BibitemOpen
  \bibfield  {author} {\bibinfo {author} {\bibfnamefont {J.}~\bibnamefont
  {Jadczak}}, \bibinfo {author} {\bibfnamefont {P.}~\bibnamefont {Plochocka}},
  \bibinfo {author} {\bibfnamefont {A.}~\bibnamefont {Mitioglu}}, \bibinfo
  {author} {\bibfnamefont {I.}~\bibnamefont {Breslavetz}}, \bibinfo {author}
  {\bibfnamefont {M.}~\bibnamefont {Royo}}, \bibinfo {author} {\bibfnamefont
  {A.}~\bibnamefont {Bertoni}}, \bibinfo {author} {\bibfnamefont
  {G.}~\bibnamefont {Goldoni}}, \bibinfo {author} {\bibfnamefont
  {T.}~\bibnamefont {Smolenski}}, \bibinfo {author} {\bibfnamefont
  {P.}~\bibnamefont {Kossacki}}, \bibinfo {author} {\bibfnamefont
  {A.}~\bibnamefont {Kretinin}}, \bibinfo {author} {\bibfnamefont
  {H.}~\bibnamefont {Shtrikman}}, \ and\ \bibinfo {author} {\bibfnamefont
  {D.~K.}\ \bibnamefont {Maude}},\ }\href@noop {} {\bibfield  {journal}
  {\bibinfo  {journal} {Nano Letters}\ }\textbf {\bibinfo {volume} {14}},\
  \bibinfo {pages} {2807} (\bibinfo {year} {2014})}\BibitemShut {NoStop}%
\bibitem [{\citenamefont {Qian}\ \emph {et~al.}(2004)\citenamefont {Qian},
  \citenamefont {Li}, \citenamefont {Grade{\v{c}}ak}, \citenamefont {Wang},
  \citenamefont {Barrelet},\ and\ \citenamefont {Lieber}}]{Qian04}%
  \BibitemOpen
  \bibfield  {author} {\bibinfo {author} {\bibfnamefont {F.}~\bibnamefont
  {Qian}}, \bibinfo {author} {\bibfnamefont {Y.}~\bibnamefont {Li}}, \bibinfo
  {author} {\bibfnamefont {S.}~\bibnamefont {Grade{\v{c}}ak}}, \bibinfo
  {author} {\bibfnamefont {D.}~\bibnamefont {Wang}}, \bibinfo {author}
  {\bibfnamefont {C.~J.}\ \bibnamefont {Barrelet}}, \ and\ \bibinfo {author}
  {\bibfnamefont {C.~M.}\ \bibnamefont {Lieber}},\ }\href@noop {} {\bibfield
  {journal} {\bibinfo  {journal} {Nano Letters}\ }\textbf {\bibinfo {volume}
  {4}},\ \bibinfo {pages} {1975} (\bibinfo {year} {2004})}\BibitemShut
  {NoStop}%
\bibitem [{\citenamefont {Qian}\ \emph {et~al.}(2005)\citenamefont {Qian},
  \citenamefont {Grade{\v{c}}ak}, \citenamefont {Li}, \citenamefont {Wen},\
  and\ \citenamefont {Lieber}}]{Qian05}%
  \BibitemOpen
  \bibfield  {author} {\bibinfo {author} {\bibfnamefont {F.}~\bibnamefont
  {Qian}}, \bibinfo {author} {\bibfnamefont {S.}~\bibnamefont
  {Grade{\v{c}}ak}}, \bibinfo {author} {\bibfnamefont {Y.}~\bibnamefont {Li}},
  \bibinfo {author} {\bibfnamefont {C.-Y.}\ \bibnamefont {Wen}}, \ and\
  \bibinfo {author} {\bibfnamefont {C.~M.}\ \bibnamefont {Lieber}},\
  }\href@noop {} {\bibfield  {journal} {\bibinfo  {journal} {Nano Letters}\
  }\textbf {\bibinfo {volume} {5}},\ \bibinfo {pages} {2287} (\bibinfo {year}
  {2005})}\BibitemShut {NoStop}%
\bibitem [{\citenamefont {Baird}\ \emph {et~al.}(2009)\citenamefont {Baird},
  \citenamefont {Ang}, \citenamefont {Low}, \citenamefont {Haegel},
  \citenamefont {Talin}, \citenamefont {Li},\ and\ \citenamefont
  {Wang}}]{Baird09}%
  \BibitemOpen
  \bibfield  {author} {\bibinfo {author} {\bibfnamefont {L.}~\bibnamefont
  {Baird}}, \bibinfo {author} {\bibfnamefont {G.}~\bibnamefont {Ang}}, \bibinfo
  {author} {\bibfnamefont {C.}~\bibnamefont {Low}}, \bibinfo {author}
  {\bibfnamefont {N.}~\bibnamefont {Haegel}}, \bibinfo {author} {\bibfnamefont
  {A.}~\bibnamefont {Talin}}, \bibinfo {author} {\bibfnamefont
  {Q.}~\bibnamefont {Li}}, \ and\ \bibinfo {author} {\bibfnamefont
  {G.}~\bibnamefont {Wang}},\ }\href@noop {} {\bibfield  {journal} {\bibinfo
  {journal} {Physica B: Condensed Matter}\ }\textbf {\bibinfo {volume} {404}},\
  \bibinfo {pages} {4933 } (\bibinfo {year} {2009})}\BibitemShut {NoStop}%
\bibitem [{\citenamefont {Heurlin}\ \emph {et~al.}(2015)\citenamefont
  {Heurlin}, \citenamefont {Stankevi{\v{c}}}, \citenamefont
  {Mickevi{\v{c}}ius}, \citenamefont {Yngman}, \citenamefont {Lindgren},
  \citenamefont {Mikkelsen}, \citenamefont {Feidenhans’l}, \citenamefont
  {Borgst{\"o}m},\ and\ \citenamefont {Samuelson}}]{Heurlin15}%
  \BibitemOpen
  \bibfield  {author} {\bibinfo {author} {\bibfnamefont {M.}~\bibnamefont
  {Heurlin}}, \bibinfo {author} {\bibfnamefont {T.}~\bibnamefont
  {Stankevi{\v{c}}}}, \bibinfo {author} {\bibfnamefont {S.}~\bibnamefont
  {Mickevi{\v{c}}ius}}, \bibinfo {author} {\bibfnamefont {S.}~\bibnamefont
  {Yngman}}, \bibinfo {author} {\bibfnamefont {D.}~\bibnamefont {Lindgren}},
  \bibinfo {author} {\bibfnamefont {A.}~\bibnamefont {Mikkelsen}}, \bibinfo
  {author} {\bibfnamefont {R.}~\bibnamefont {Feidenhans’l}}, \bibinfo
  {author} {\bibfnamefont {M.~T.}\ \bibnamefont {Borgst{\"o}m}}, \ and\
  \bibinfo {author} {\bibfnamefont {L.}~\bibnamefont {Samuelson}},\ }\href@noop
  {} {\bibfield  {journal} {\bibinfo  {journal} {Nano Letters}\ }\textbf
  {\bibinfo {volume} {15}},\ \bibinfo {pages} {2462} (\bibinfo {year}
  {2015})}\BibitemShut {NoStop}%
\bibitem [{\citenamefont {Dong}\ \emph {et~al.}(2009)\citenamefont {Dong},
  \citenamefont {Tian}, \citenamefont {Kempa},\ and\ \citenamefont
  {Lieber}}]{Dong09}%
  \BibitemOpen
  \bibfield  {author} {\bibinfo {author} {\bibfnamefont {Y.}~\bibnamefont
  {Dong}}, \bibinfo {author} {\bibfnamefont {B.}~\bibnamefont {Tian}}, \bibinfo
  {author} {\bibfnamefont {T.~J.}\ \bibnamefont {Kempa}}, \ and\ \bibinfo
  {author} {\bibfnamefont {C.~M.}\ \bibnamefont {Lieber}},\ }\href@noop {}
  {\bibfield  {journal} {\bibinfo  {journal} {Nano Letters}\ }\textbf {\bibinfo
  {volume} {9}},\ \bibinfo {pages} {2183} (\bibinfo {year} {2009})}\BibitemShut
  {NoStop}%
\bibitem [{\citenamefont {Yuan}\ \emph {et~al.}(2015)\citenamefont {Yuan},
  \citenamefont {Caroff}, \citenamefont {Wang}, \citenamefont {Guo},
  \citenamefont {Wang}, \citenamefont {Jackson}, \citenamefont {Smith},
  \citenamefont {Tan},\ and\ \citenamefont {Jagadish}}]{Yuan15}%
  \BibitemOpen
  \bibfield  {author} {\bibinfo {author} {\bibfnamefont {X.}~\bibnamefont
  {Yuan}}, \bibinfo {author} {\bibfnamefont {P.}~\bibnamefont {Caroff}},
  \bibinfo {author} {\bibfnamefont {F.}~\bibnamefont {Wang}}, \bibinfo {author}
  {\bibfnamefont {Y.}~\bibnamefont {Guo}}, \bibinfo {author} {\bibfnamefont
  {Y.}~\bibnamefont {Wang}}, \bibinfo {author} {\bibfnamefont {H.~E.}\
  \bibnamefont {Jackson}}, \bibinfo {author} {\bibfnamefont {L.~M.}\
  \bibnamefont {Smith}}, \bibinfo {author} {\bibfnamefont {H.~H.}\ \bibnamefont
  {Tan}}, \ and\ \bibinfo {author} {\bibfnamefont {C.}~\bibnamefont
  {Jagadish}},\ }\href@noop {} {\bibfield  {journal} {\bibinfo  {journal} {Adv.
  Funct. Mater.}\ }\textbf {\bibinfo {volume} {25}},\ \bibinfo {pages} {5300}
  (\bibinfo {year} {2015})}\BibitemShut {NoStop}%
\bibitem [{\citenamefont {G\"{o}ransson}\ \emph {et~al.}(2019)\citenamefont
  {G\"{o}ransson}, \citenamefont {Heurlin}, \citenamefont {Dalelkhan},
  \citenamefont {Abay}, \citenamefont {Messing}, \citenamefont {Maisi},
  \citenamefont {Borgström},\ and\ \citenamefont {Xu}}]{Goransson19}%
  \BibitemOpen
  \bibfield  {author} {\bibinfo {author} {\bibfnamefont {D.~J.~O.}\
  \bibnamefont {G\"{o}ransson}}, \bibinfo {author} {\bibfnamefont
  {M.}~\bibnamefont {Heurlin}}, \bibinfo {author} {\bibfnamefont
  {B.}~\bibnamefont {Dalelkhan}}, \bibinfo {author} {\bibfnamefont
  {S.}~\bibnamefont {Abay}}, \bibinfo {author} {\bibfnamefont {M.~E.}\
  \bibnamefont {Messing}}, \bibinfo {author} {\bibfnamefont {V.~F.}\
  \bibnamefont {Maisi}}, \bibinfo {author} {\bibfnamefont {M.~T.}\ \bibnamefont
  {Borgström}}, \ and\ \bibinfo {author} {\bibfnamefont {H.~Q.}\ \bibnamefont
  {Xu}},\ }\href {\doibase 10.1063/1.5084222} {\bibfield  {journal} {\bibinfo
  {journal} {Applied Physics Letters}\ }\textbf {\bibinfo {volume} {114}},\
  \bibinfo {pages} {053108} (\bibinfo {year} {2019})}\BibitemShut {NoStop}%
\bibitem [{\citenamefont {Rieger}\ \emph {et~al.}(2015)\citenamefont {Rieger},
  \citenamefont {Grutzmacher},\ and\ \citenamefont {Lepsa}}]{Rieger15}%
  \BibitemOpen
  \bibfield  {author} {\bibinfo {author} {\bibfnamefont {T.}~\bibnamefont
  {Rieger}}, \bibinfo {author} {\bibfnamefont {D.}~\bibnamefont {Grutzmacher}},
  \ and\ \bibinfo {author} {\bibfnamefont {M.~I.}\ \bibnamefont {Lepsa}},\
  }\href@noop {} {\bibfield  {journal} {\bibinfo  {journal} {Nanoscale}\
  }\textbf {\bibinfo {volume} {7}},\ \bibinfo {pages} {356} (\bibinfo {year}
  {2015})}\BibitemShut {NoStop}%
\bibitem [{\citenamefont {Kim}\ and\ \citenamefont {No}(2017)}]{Kim2017}%
  \BibitemOpen
  \bibfield  {author} {\bibinfo {author} {\bibfnamefont {K.-H.}\ \bibnamefont
  {Kim}}\ and\ \bibinfo {author} {\bibfnamefont {Y.-S.}\ \bibnamefont {No}},\
  }\href {\doibase 10.1186/s40580-017-0128-8} {\bibfield  {journal} {\bibinfo
  {journal} {Nano Convergence}\ }\textbf {\bibinfo {volume} {4}},\ \bibinfo
  {pages} {32} (\bibinfo {year} {2017})}\BibitemShut {NoStop}%
\bibitem [{\citenamefont {Ferrari}\ \emph {et~al.}(2009)\citenamefont
  {Ferrari}, \citenamefont {Goldoni}, \citenamefont {Bertoni}, \citenamefont
  {Cuoghi},\ and\ \citenamefont {Molinari}}]{Ferrari09b}%
  \BibitemOpen
  \bibfield  {author} {\bibinfo {author} {\bibfnamefont {G.}~\bibnamefont
  {Ferrari}}, \bibinfo {author} {\bibfnamefont {G.}~\bibnamefont {Goldoni}},
  \bibinfo {author} {\bibfnamefont {A.}~\bibnamefont {Bertoni}}, \bibinfo
  {author} {\bibfnamefont {G.}~\bibnamefont {Cuoghi}}, \ and\ \bibinfo {author}
  {\bibfnamefont {E.}~\bibnamefont {Molinari}},\ }\href@noop {} {\bibfield
  {journal} {\bibinfo  {journal} {Nano Letters}\ }\textbf {\bibinfo {volume}
  {9}},\ \bibinfo {pages} {1631} (\bibinfo {year} {2009})}\BibitemShut
  {NoStop}%
\bibitem [{\citenamefont {Wong}\ \emph {et~al.}(2011)\citenamefont {Wong},
  \citenamefont {Léonard}, \citenamefont {Li},\ and\ \citenamefont
  {Wang}}]{Wong11}%
  \BibitemOpen
  \bibfield  {author} {\bibinfo {author} {\bibfnamefont {B.~M.}\ \bibnamefont
  {Wong}}, \bibinfo {author} {\bibfnamefont {F.}~\bibnamefont {Léonard}},
  \bibinfo {author} {\bibfnamefont {Q.}~\bibnamefont {Li}}, \ and\ \bibinfo
  {author} {\bibfnamefont {G.~T.}\ \bibnamefont {Wang}},\ }\href@noop {}
  {\bibfield  {journal} {\bibinfo  {journal} {Nano Letters}\ }\textbf {\bibinfo
  {volume} {11}},\ \bibinfo {pages} {3074} (\bibinfo {year}
  {2011})}\BibitemShut {NoStop}%
\bibitem [{\citenamefont {Sitek}\ \emph {et~al.}(2015)\citenamefont {Sitek},
  \citenamefont {Serra}, \citenamefont {Gudmundsson},\ and\ \citenamefont
  {Manolescu}}]{Sitek15}%
  \BibitemOpen
  \bibfield  {author} {\bibinfo {author} {\bibfnamefont {A.}~\bibnamefont
  {Sitek}}, \bibinfo {author} {\bibfnamefont {L.}~\bibnamefont {Serra}},
  \bibinfo {author} {\bibfnamefont {V.}~\bibnamefont {Gudmundsson}}, \ and\
  \bibinfo {author} {\bibfnamefont {A.}~\bibnamefont {Manolescu}},\ }\href@noop
  {} {\bibfield  {journal} {\bibinfo  {journal} {Phys. Rev. B}\ }\textbf
  {\bibinfo {volume} {91}},\ \bibinfo {pages} {235429} (\bibinfo {year}
  {2015})}\BibitemShut {NoStop}%
\bibitem [{\citenamefont {Sitek}\ \emph {et~al.}(2016)\citenamefont {Sitek},
  \citenamefont {Thorgilsson}, \citenamefont {Gudmundsson},\ and\ \citenamefont
  {Manolescu}}]{Sitek16}%
  \BibitemOpen
  \bibfield  {author} {\bibinfo {author} {\bibfnamefont {A.}~\bibnamefont
  {Sitek}}, \bibinfo {author} {\bibfnamefont {G.}~\bibnamefont {Thorgilsson}},
  \bibinfo {author} {\bibfnamefont {V.}~\bibnamefont {Gudmundsson}}, \ and\
  \bibinfo {author} {\bibfnamefont {A.}~\bibnamefont {Manolescu}},\ }\href@noop
  {} {\bibfield  {journal} {\bibinfo  {journal} {Nanotechnology}\ }\textbf
  {\bibinfo {volume} {27}},\ \bibinfo {pages} {225202} (\bibinfo {year}
  {2016})}\BibitemShut {NoStop}%
\bibitem [{\citenamefont {Royo}\ \emph {et~al.}(2017)\citenamefont {Royo},
  \citenamefont {Luca}, \citenamefont {Rurali},\ and\ \citenamefont
  {Zardo}}]{Royo_2017}%
  \BibitemOpen
  \bibfield  {author} {\bibinfo {author} {\bibfnamefont {M.}~\bibnamefont
  {Royo}}, \bibinfo {author} {\bibfnamefont {M.~D.}\ \bibnamefont {Luca}},
  \bibinfo {author} {\bibfnamefont {R.}~\bibnamefont {Rurali}}, \ and\ \bibinfo
  {author} {\bibfnamefont {I.}~\bibnamefont {Zardo}},\ }\href {\doibase
  10.1088/1361-6463/aa5d8e} {\bibfield  {journal} {\bibinfo  {journal} {Journal
  of Physics D: Applied Physics}\ }\textbf {\bibinfo {volume} {50}},\ \bibinfo
  {pages} {143001} (\bibinfo {year} {2017})}\BibitemShut {NoStop}%
\bibitem [{\citenamefont {Shen}\ and\ \citenamefont {Chen}(2009)}]{Shen2009}%
  \BibitemOpen
  \bibfield  {author} {\bibinfo {author} {\bibfnamefont {G.}~\bibnamefont
  {Shen}}\ and\ \bibinfo {author} {\bibfnamefont {D.}~\bibnamefont {Chen}},\
  }\href@noop {} {\bibfield  {journal} {\bibinfo  {journal} {Nanoscale Research
  Letters}\ }\textbf {\bibinfo {volume} {4}},\ \bibinfo {pages} {779} (\bibinfo
  {year} {2009})}\BibitemShut {NoStop}%
\bibitem [{\citenamefont {Koblm{\"u}ller}\ \emph {et~al.}(2017)\citenamefont
  {Koblm{\"u}ller}, \citenamefont {Mayer}, \citenamefont {Stettner},
  \citenamefont {Abstreiter},\ and\ \citenamefont {Finley}}]{Koblm_ller_2017}%
  \BibitemOpen
  \bibfield  {author} {\bibinfo {author} {\bibfnamefont {G.}~\bibnamefont
  {Koblm{\"u}ller}}, \bibinfo {author} {\bibfnamefont {B.}~\bibnamefont
  {Mayer}}, \bibinfo {author} {\bibfnamefont {T.}~\bibnamefont {Stettner}},
  \bibinfo {author} {\bibfnamefont {G.}~\bibnamefont {Abstreiter}}, \ and\
  \bibinfo {author} {\bibfnamefont {J.~J.}\ \bibnamefont {Finley}},\
  }\href@noop {} {\bibfield  {journal} {\bibinfo  {journal} {Semiconductor
  Science and Technology}\ }\textbf {\bibinfo {volume} {32}},\ \bibinfo {pages}
  {053001} (\bibinfo {year} {2017})}\BibitemShut {NoStop}%
\bibitem [{\citenamefont {Florica}\ \emph {et~al.}(2019)\citenamefont
  {Florica}, \citenamefont {Costas}, \citenamefont {Preda}, \citenamefont
  {Beregoi}, \citenamefont {Kuncser}, \citenamefont {Apostol}, \citenamefont
  {Popa}, \citenamefont {Socol}, \citenamefont {Diculescu},\ and\ \citenamefont
  {Enculescu}}]{Florica2019}%
  \BibitemOpen
  \bibfield  {author} {\bibinfo {author} {\bibfnamefont {C.}~\bibnamefont
  {Florica}}, \bibinfo {author} {\bibfnamefont {A.}~\bibnamefont {Costas}},
  \bibinfo {author} {\bibfnamefont {N.}~\bibnamefont {Preda}}, \bibinfo
  {author} {\bibfnamefont {M.}~\bibnamefont {Beregoi}}, \bibinfo {author}
  {\bibfnamefont {A.}~\bibnamefont {Kuncser}}, \bibinfo {author} {\bibfnamefont
  {N.}~\bibnamefont {Apostol}}, \bibinfo {author} {\bibfnamefont
  {C.}~\bibnamefont {Popa}}, \bibinfo {author} {\bibfnamefont {G.}~\bibnamefont
  {Socol}}, \bibinfo {author} {\bibfnamefont {V.}~\bibnamefont {Diculescu}}, \
  and\ \bibinfo {author} {\bibfnamefont {I.}~\bibnamefont {Enculescu}},\ }\href
  {\doibase 10.1038/s41598-019-53873-0} {\bibfield  {journal} {\bibinfo
  {journal} {Scientific Reports}\ }\textbf {\bibinfo {volume} {9}},\ \bibinfo
  {pages} {17268} (\bibinfo {year} {2019})}\BibitemShut {NoStop}%
\bibitem [{\citenamefont {Hassan}\ \emph {et~al.}(2019)\citenamefont {Hassan},
  \citenamefont {Johar}, \citenamefont {Waseem}, \citenamefont {Bagal},
  \citenamefont {Ha},\ and\ \citenamefont {Ryu}}]{Hassan:19}%
  \BibitemOpen
  \bibfield  {author} {\bibinfo {author} {\bibfnamefont {M.~A.}\ \bibnamefont
  {Hassan}}, \bibinfo {author} {\bibfnamefont {M.~A.}\ \bibnamefont {Johar}},
  \bibinfo {author} {\bibfnamefont {A.}~\bibnamefont {Waseem}}, \bibinfo
  {author} {\bibfnamefont {I.~V.}\ \bibnamefont {Bagal}}, \bibinfo {author}
  {\bibfnamefont {J.-S.}\ \bibnamefont {Ha}}, \ and\ \bibinfo {author}
  {\bibfnamefont {S.-W.}\ \bibnamefont {Ryu}},\ }\href {\doibase
  10.1364/OE.27.00A184} {\bibfield  {journal} {\bibinfo  {journal} {Opt.
  Express}\ }\textbf {\bibinfo {volume} {27}},\ \bibinfo {pages} {A184}
  (\bibinfo {year} {2019})}\BibitemShut {NoStop}%
\bibitem [{\citenamefont {Oener}\ \emph {et~al.}(2015)\citenamefont {Oener},
  \citenamefont {Mann}, \citenamefont {Sciacca}, \citenamefont {Sfiligoj},
  \citenamefont {Hoang},\ and\ \citenamefont {Garnett}}]{core-shell_photo}%
  \BibitemOpen
  \bibfield  {author} {\bibinfo {author} {\bibfnamefont {S.~Z.}\ \bibnamefont
  {Oener}}, \bibinfo {author} {\bibfnamefont {S.~A.}\ \bibnamefont {Mann}},
  \bibinfo {author} {\bibfnamefont {B.}~\bibnamefont {Sciacca}}, \bibinfo
  {author} {\bibfnamefont {C.}~\bibnamefont {Sfiligoj}}, \bibinfo {author}
  {\bibfnamefont {J.}~\bibnamefont {Hoang}}, \ and\ \bibinfo {author}
  {\bibfnamefont {E.~C.}\ \bibnamefont {Garnett}},\ }\href {\doibase
  10.1063/1.4905652} {\bibfield  {journal} {\bibinfo  {journal} {Applied
  Physics Letters}\ }\textbf {\bibinfo {volume} {106}},\ \bibinfo {pages}
  {023501} (\bibinfo {year} {2015})},\ \Eprint
  {http://arxiv.org/abs/https://doi.org/10.1063/1.4905652}
  {https://doi.org/10.1063/1.4905652} \BibitemShut {NoStop}%
\bibitem [{\citenamefont {Manolescu}\ \emph {et~al.}(2017)\citenamefont
  {Manolescu}, \citenamefont {Sitek}, \citenamefont {Osca}, \citenamefont
  {Serra}, \citenamefont {Gudmundsson},\ and\ \citenamefont
  {Stanescu}}]{Andrei}%
  \BibitemOpen
  \bibfield  {author} {\bibinfo {author} {\bibfnamefont {A.}~\bibnamefont
  {Manolescu}}, \bibinfo {author} {\bibfnamefont {A.}~\bibnamefont {Sitek}},
  \bibinfo {author} {\bibfnamefont {J.}~\bibnamefont {Osca}}, \bibinfo {author}
  {\bibfnamefont {L.}~\bibnamefont {Serra}}, \bibinfo {author} {\bibfnamefont
  {V.}~\bibnamefont {Gudmundsson}}, \ and\ \bibinfo {author} {\bibfnamefont
  {T.~D.}\ \bibnamefont {Stanescu}},\ }\href@noop {} {\bibfield  {journal}
  {\bibinfo  {journal} {Phys. Rev. B}\ }\textbf {\bibinfo {volume} {96}},\
  \bibinfo {pages} {125435} (\bibinfo {year} {2017})}\BibitemShut {NoStop}%
\bibitem [{\citenamefont {Klausen}\ \emph {et~al.}(2020)\citenamefont
  {Klausen}, \citenamefont {Sitek}, \citenamefont {Erlingsson},\ and\
  \citenamefont {Manolescu}}]{Ottar_Klausen_2020}%
  \BibitemOpen
  \bibfield  {author} {\bibinfo {author} {\bibfnamefont {K.~O.}\ \bibnamefont
  {Klausen}}, \bibinfo {author} {\bibfnamefont {A.}~\bibnamefont {Sitek}},
  \bibinfo {author} {\bibfnamefont {S.~I.}\ \bibnamefont {Erlingsson}}, \ and\
  \bibinfo {author} {\bibfnamefont {A.}~\bibnamefont {Manolescu}},\ }\href
  {\doibase 10.1088/1361-6528/ab932e} {\bibfield  {journal} {\bibinfo
  {journal} {Nanotechnology}\ }\textbf {\bibinfo {volume} {31}},\ \bibinfo
  {pages} {354001} (\bibinfo {year} {2020})}\BibitemShut {NoStop}%
\bibitem [{\citenamefont {McMillan}(1968)}]{MacMillan_NS_Theory}%
  \BibitemOpen
  \bibfield  {author} {\bibinfo {author} {\bibfnamefont {W.~L.}\ \bibnamefont
  {McMillan}},\ }\href {\doibase 10.1103/PhysRev.175.559} {\bibfield  {journal}
  {\bibinfo  {journal} {Phys. Rev.}\ }\textbf {\bibinfo {volume} {175}},\
  \bibinfo {pages} {559} (\bibinfo {year} {1968})}\BibitemShut {NoStop}%
\bibitem [{\citenamefont {Gor'kov}(1958)}]{Gorkov_58}%
  \BibitemOpen
  \bibfield  {author} {\bibinfo {author} {\bibfnamefont {L.}~\bibnamefont
  {Gor'kov}},\ }\href@noop {} {\bibfield  {journal} {\bibinfo  {journal} {Sov.
  Phys. - JETP (Engl. Transl.); (United States)}\ }\textbf {\bibinfo {volume}
  {7:3}} (\bibinfo {year} {1958})}\BibitemShut {NoStop}%
\bibitem [{\citenamefont {Fetter}\ and\ \citenamefont
  {Walecka}(2003)}]{FetterW}%
  \BibitemOpen
  \bibfield  {author} {\bibinfo {author} {\bibfnamefont {A.~L.}\ \bibnamefont
  {Fetter}}\ and\ \bibinfo {author} {\bibfnamefont {J.~D.}\ \bibnamefont
  {Walecka}},\ }\href@noop {} {\emph {\bibinfo {title} {Quantum Theory of
  Many-Particle Systems}}}\ (\bibinfo  {publisher} {New York: Dover.},\
  \bibinfo {year} {2003})\BibitemShut {NoStop}%
\bibitem [{\citenamefont {Zheng}\ and\ \citenamefont
  {Walmsley}(2005)}]{Zheng_2005}%
  \BibitemOpen
  \bibfield  {author} {\bibinfo {author} {\bibfnamefont {X.~H.}\ \bibnamefont
  {Zheng}}\ and\ \bibinfo {author} {\bibfnamefont {D.~G.}\ \bibnamefont
  {Walmsley}},\ }\href {\doibase 10.1103/PhysRevB.71.134512} {\bibfield
  {journal} {\bibinfo  {journal} {Phys. Rev. B}\ }\textbf {\bibinfo {volume}
  {71}},\ \bibinfo {pages} {134512} (\bibinfo {year} {2005})}\BibitemShut
  {NoStop}%
\bibitem [{\citenamefont {Nambu}(1960)}]{Nambu_1960}%
  \BibitemOpen
  \bibfield  {author} {\bibinfo {author} {\bibfnamefont {Y.}~\bibnamefont
  {Nambu}},\ }\href {\doibase 10.1103/PhysRev.117.648} {\bibfield  {journal}
  {\bibinfo  {journal} {Phys. Rev.}\ }\textbf {\bibinfo {volume} {117}},\
  \bibinfo {pages} {648} (\bibinfo {year} {1960})}\BibitemShut {NoStop}%
\bibitem [{\citenamefont {Bruus}\ and\ \citenamefont
  {Flensberg}(2004)}]{BruusFlensberg}%
  \BibitemOpen
  \bibfield  {author} {\bibinfo {author} {\bibfnamefont {H.}~\bibnamefont
  {Bruus}}\ and\ \bibinfo {author} {\bibfnamefont {K.}~\bibnamefont
  {Flensberg}},\ }\href@noop {} {\emph {\bibinfo {title} {Many-body quantum
  theory in condensed matter physics - an introduction}}}\ (\bibinfo
  {publisher} {Oxford University Press},\ \bibinfo {address} {United States},\
  \bibinfo {year} {2004})\BibitemShut {NoStop}%
\bibitem [{\citenamefont {Deutscher}\ and\ \citenamefont
  {de~Gennes}(1969)}]{DeGennesDeutcher}%
  \BibitemOpen
  \bibfield  {author} {\bibinfo {author} {\bibfnamefont {G.}~\bibnamefont
  {Deutscher}}\ and\ \bibinfo {author} {\bibfnamefont {P.}~\bibnamefont
  {de~Gennes}},\ }\href@noop {} {\bibfield  {journal} {\bibinfo  {journal} {pp
  1005-34 of Superconductivity. Vols. 1 and 2. Parks, R. D. (ed.). New York,
  Marcel Dekker, Inc., 1969.}\ } (\bibinfo {year} {1969})}\BibitemShut
  {NoStop}%
\bibitem [{\citenamefont {Falk}(1963)}]{Falk}%
  \BibitemOpen
  \bibfield  {author} {\bibinfo {author} {\bibfnamefont {D.~S.}\ \bibnamefont
  {Falk}},\ }\href {\doibase 10.1103/PhysRev.132.1576} {\bibfield  {journal}
  {\bibinfo  {journal} {Phys. Rev.}\ }\textbf {\bibinfo {volume} {132}},\
  \bibinfo {pages} {1576} (\bibinfo {year} {1963})}\BibitemShut {NoStop}%
\bibitem [{\citenamefont {Tinkham}(2003)}]{Tinkham}%
  \BibitemOpen
  \bibfield  {author} {\bibinfo {author} {\bibfnamefont {M.}~\bibnamefont
  {Tinkham}},\ }\href@noop {} {\emph {\bibinfo {title} {Introduction to
  Superconductivity}}}\ (\bibinfo  {publisher} {Dover Publications, Inc.
  Mineola, New York.},\ \bibinfo {year} {2003})\BibitemShut {NoStop}%
\bibitem [{\citenamefont {Sch{\"a}pers}(2001)}]{Schapers}%
  \BibitemOpen
  \bibfield  {author} {\bibinfo {author} {\bibfnamefont {T.}~\bibnamefont
  {Sch{\"a}pers}},\ }\href@noop {} {\emph {\bibinfo {title}
  {Superconductor/Semiconductor Junctions}}}\ (\bibinfo  {publisher}
  {Springer-Verlag Berlin Heidelberg},\ \bibinfo {year} {2001})\BibitemShut
  {NoStop}%
\bibitem [{\citenamefont {{van Houten}}\ and\ \citenamefont
  {Beenakker}(1991)}]{Beenakker91}%
  \BibitemOpen
  \bibfield  {author} {\bibinfo {author} {\bibfnamefont {H.}~\bibnamefont {{van
  Houten}}}\ and\ \bibinfo {author} {\bibfnamefont {C.}~\bibnamefont
  {Beenakker}},\ }\href {\doibase https://doi.org/10.1016/0921-4526(91)90712-N}
  {\bibfield  {journal} {\bibinfo  {journal} {Physica B: Condensed Matter}\
  }\textbf {\bibinfo {volume} {175}},\ \bibinfo {pages} {187} (\bibinfo {year}
  {1991})},\ \bibinfo {note} {analogies in Optics and
  Micro-Electronics}\BibitemShut {NoStop}%
\bibitem [{\citenamefont {Martin}\ and\ \citenamefont
  {Annett}(1999)}]{MARTIN99}%
  \BibitemOpen
  \bibfield  {author} {\bibinfo {author} {\bibfnamefont {A.}~\bibnamefont
  {Martin}}\ and\ \bibinfo {author} {\bibfnamefont {J.~F.}\ \bibnamefont
  {Annett}},\ }\href {\doibase https://doi.org/10.1006/spmi.1999.0709}
  {\bibfield  {journal} {\bibinfo  {journal} {Superlattices and
  Microstructures}\ }\textbf {\bibinfo {volume} {25}},\ \bibinfo {pages} {1019}
  (\bibinfo {year} {1999})}\BibitemShut {NoStop}%
\bibitem [{\citenamefont {Aizawa}\ and\ \citenamefont
  {Kuroki}(2018)}]{Aizawa_2018}%
  \BibitemOpen
  \bibfield  {author} {\bibinfo {author} {\bibfnamefont {H.}~\bibnamefont
  {Aizawa}}\ and\ \bibinfo {author} {\bibfnamefont {K.}~\bibnamefont
  {Kuroki}},\ }\href {\doibase 10.1103/PhysRevB.97.104507} {\bibfield
  {journal} {\bibinfo  {journal} {Phys. Rev. B}\ }\textbf {\bibinfo {volume}
  {97}},\ \bibinfo {pages} {104507} (\bibinfo {year} {2018})}\BibitemShut
  {NoStop}%
\bibitem [{\citenamefont {Gu\'eron}\ \emph {et~al.}(1996)\citenamefont
  {Gu\'eron}, \citenamefont {Pothier}, \citenamefont {Birge}, \citenamefont
  {Esteve},\ and\ \citenamefont {Devoret}}]{Devoret_96}%
  \BibitemOpen
  \bibfield  {author} {\bibinfo {author} {\bibfnamefont {S.}~\bibnamefont
  {Gu\'eron}}, \bibinfo {author} {\bibfnamefont {H.}~\bibnamefont {Pothier}},
  \bibinfo {author} {\bibfnamefont {N.~O.}\ \bibnamefont {Birge}}, \bibinfo
  {author} {\bibfnamefont {D.}~\bibnamefont {Esteve}}, \ and\ \bibinfo {author}
  {\bibfnamefont {M.~H.}\ \bibnamefont {Devoret}},\ }\href {\doibase
  10.1103/PhysRevLett.77.3025} {\bibfield  {journal} {\bibinfo  {journal}
  {Phys. Rev. Lett.}\ }\textbf {\bibinfo {volume} {77}},\ \bibinfo {pages}
  {3025} (\bibinfo {year} {1996})}\BibitemShut {NoStop}%
\bibitem [{\citenamefont {Usadel}(1970)}]{Usadel_70}%
  \BibitemOpen
  \bibfield  {author} {\bibinfo {author} {\bibfnamefont {K.~D.}\ \bibnamefont
  {Usadel}},\ }\href {\doibase 10.1103/PhysRevLett.25.507} {\bibfield
  {journal} {\bibinfo  {journal} {Phys. Rev. Lett.}\ }\textbf {\bibinfo
  {volume} {25}},\ \bibinfo {pages} {507} (\bibinfo {year} {1970})}\BibitemShut
  {NoStop}%
\bibitem [{\citenamefont {Eilenberger}(1968)}]{Eilenberger_68}%
  \BibitemOpen
  \bibfield  {author} {\bibinfo {author} {\bibfnamefont {G.}~\bibnamefont
  {Eilenberger}},\ }\href {\doibase 10.1007/BF01379803} {\bibfield  {journal}
  {\bibinfo  {journal} {Zeitschrift f{\"u}r Physik A Hadrons and nuclei}\
  }\textbf {\bibinfo {volume} {214}},\ \bibinfo {pages} {195} (\bibinfo {year}
  {1968})}\BibitemShut {NoStop}%
\bibitem [{\citenamefont {{Larkin}}\ and\ \citenamefont
  {{Ovchinnikov}}(1969)}]{LarkinOvchinnikov_69}%
  \BibitemOpen
  \bibfield  {author} {\bibinfo {author} {\bibfnamefont {A.~I.}\ \bibnamefont
  {{Larkin}}}\ and\ \bibinfo {author} {\bibfnamefont {Y.~N.}\ \bibnamefont
  {{Ovchinnikov}}},\ }\href@noop {} {\bibfield  {journal} {\bibinfo  {journal}
  {Soviet Journal of Experimental and Theoretical Physics}\ }\textbf {\bibinfo
  {volume} {28}},\ \bibinfo {pages} {1200} (\bibinfo {year}
  {1969})}\BibitemShut {NoStop}%
\bibitem [{\citenamefont {Ginzburg}\ and\ \citenamefont {Landau}(1950)}]{GL}%
  \BibitemOpen
  \bibfield  {author} {\bibinfo {author} {\bibfnamefont {V.~L.}\ \bibnamefont
  {Ginzburg}}\ and\ \bibinfo {author} {\bibfnamefont {L.~D.}\ \bibnamefont
  {Landau}},\ }\href@noop {} {\bibfield  {journal} {\bibinfo  {journal} {Zh.
  Eksp. Teor. Fiz.}\ }\textbf {\bibinfo {volume} {20}},\ \bibinfo {pages}
  {1064} (\bibinfo {year} {1950})}\BibitemShut {NoStop}%
\bibitem [{\citenamefont {Gor'kov}(1959)}]{Gorkov_59}%
  \BibitemOpen
  \bibfield  {author} {\bibinfo {author} {\bibfnamefont {L.~P.}\ \bibnamefont
  {Gor'kov}},\ }\href {https://www.osti.gov/biblio/7264935} {\bibfield
  {journal} {\bibinfo  {journal} {Sov. Phys. - JETP (Engl. Transl.); (United
  States)}\ }\textbf {\bibinfo {volume} {9:6}} (\bibinfo {year}
  {1959})}\BibitemShut {NoStop}%
\bibitem [{\citenamefont {Ridderbos}\ \emph {et~al.}(2020)\citenamefont
  {Ridderbos}, \citenamefont {Brauns}, \citenamefont {de~Vries}, \citenamefont
  {Shen}, \citenamefont {Li}, \citenamefont {K{\"o}lling}, \citenamefont
  {Verheijen}, \citenamefont {Brinkman}, \citenamefont {van~der Wiel},
  \citenamefont {Bakkers},\ and\ \citenamefont {Zwanenburg}}]{Ridderbos2020}%
  \BibitemOpen
  \bibfield  {author} {\bibinfo {author} {\bibfnamefont {J.}~\bibnamefont
  {Ridderbos}}, \bibinfo {author} {\bibfnamefont {M.}~\bibnamefont {Brauns}},
  \bibinfo {author} {\bibfnamefont {F.~K.}\ \bibnamefont {de~Vries}}, \bibinfo
  {author} {\bibfnamefont {J.}~\bibnamefont {Shen}}, \bibinfo {author}
  {\bibfnamefont {A.}~\bibnamefont {Li}}, \bibinfo {author} {\bibfnamefont
  {S.}~\bibnamefont {K{\"o}lling}}, \bibinfo {author} {\bibfnamefont {M.~A.}\
  \bibnamefont {Verheijen}}, \bibinfo {author} {\bibfnamefont {A.}~\bibnamefont
  {Brinkman}}, \bibinfo {author} {\bibfnamefont {W.~G.}\ \bibnamefont {van~der
  Wiel}}, \bibinfo {author} {\bibfnamefont {E.~P. A.~M.}\ \bibnamefont
  {Bakkers}}, \ and\ \bibinfo {author} {\bibfnamefont {F.~A.}\ \bibnamefont
  {Zwanenburg}},\ }\href {\doibase 10.1021/acs.nanolett.9b03438} {\bibfield
  {journal} {\bibinfo  {journal} {Nano Letters}\ }\textbf {\bibinfo {volume}
  {20}},\ \bibinfo {pages} {122} (\bibinfo {year} {2020})}\BibitemShut
  {NoStop}%
\bibitem [{\citenamefont {Reeg}\ \emph {et~al.}(2017)\citenamefont {Reeg},
  \citenamefont {Loss},\ and\ \citenamefont {Klinovaja}}]{Reeg_2017}%
  \BibitemOpen
  \bibfield  {author} {\bibinfo {author} {\bibfnamefont {C.}~\bibnamefont
  {Reeg}}, \bibinfo {author} {\bibfnamefont {D.}~\bibnamefont {Loss}}, \ and\
  \bibinfo {author} {\bibfnamefont {J.}~\bibnamefont {Klinovaja}},\ }\href
  {\doibase 10.1103/PhysRevB.96.125426} {\bibfield  {journal} {\bibinfo
  {journal} {Phys. Rev. B}\ }\textbf {\bibinfo {volume} {96}},\ \bibinfo
  {pages} {125426} (\bibinfo {year} {2017})}\BibitemShut {NoStop}%
\bibitem [{\citenamefont {Cole}\ \emph {et~al.}(2015)\citenamefont {Cole},
  \citenamefont {Das~Sarma},\ and\ \citenamefont {Stanescu}}]{Cole15}%
  \BibitemOpen
  \bibfield  {author} {\bibinfo {author} {\bibfnamefont {W.~S.}\ \bibnamefont
  {Cole}}, \bibinfo {author} {\bibfnamefont {S.}~\bibnamefont {Das~Sarma}}, \
  and\ \bibinfo {author} {\bibfnamefont {T.~D.}\ \bibnamefont {Stanescu}},\
  }\href {\doibase 10.1103/PhysRevB.92.174511} {\bibfield  {journal} {\bibinfo
  {journal} {Phys. Rev. B}\ }\textbf {\bibinfo {volume} {92}},\ \bibinfo
  {pages} {174511} (\bibinfo {year} {2015})}\BibitemShut {NoStop}%
\bibitem [{\citenamefont {Klinovaja}\ and\ \citenamefont
  {Loss}(2012)}]{Loss_12}%
  \BibitemOpen
  \bibfield  {author} {\bibinfo {author} {\bibfnamefont {J.}~\bibnamefont
  {Klinovaja}}\ and\ \bibinfo {author} {\bibfnamefont {D.}~\bibnamefont
  {Loss}},\ }\href {\doibase 10.1103/PhysRevB.86.085408} {\bibfield  {journal}
  {\bibinfo  {journal} {Phys. Rev. B}\ }\textbf {\bibinfo {volume} {86}},\
  \bibinfo {pages} {085408} (\bibinfo {year} {2012})}\BibitemShut {NoStop}%
\bibitem [{\citenamefont {Osca}\ and\ \citenamefont
  {Serra}(2013)}]{Osca_delta}%
  \BibitemOpen
  \bibfield  {author} {\bibinfo {author} {\bibfnamefont {J.}~\bibnamefont
  {Osca}}\ and\ \bibinfo {author} {\bibfnamefont {L.}~\bibnamefont {Serra}},\
  }\href {\doibase 10.1103/PhysRevB.88.144512} {\bibfield  {journal} {\bibinfo
  {journal} {Phys. Rev. B}\ }\textbf {\bibinfo {volume} {88}},\ \bibinfo
  {pages} {144512} (\bibinfo {year} {2013})}\BibitemShut {NoStop}%
\bibitem [{\citenamefont {Livanas}\ \emph {et~al.}(2021)\citenamefont
  {Livanas}, \citenamefont {Vanas},\ and\ \citenamefont
  {Varelogiannis}}]{Livanas_2021}%
  \BibitemOpen
  \bibfield  {author} {\bibinfo {author} {\bibfnamefont {G.}~\bibnamefont
  {Livanas}}, \bibinfo {author} {\bibfnamefont {N.}~\bibnamefont {Vanas}}, \
  and\ \bibinfo {author} {\bibfnamefont {G.}~\bibnamefont {Varelogiannis}},\
  }\href {\doibase 10.3390/condmat6040044} {\bibfield  {journal} {\bibinfo
  {journal} {Condensed Matter}\ }\textbf {\bibinfo {volume} {6}},\ \bibinfo
  {pages} {44} (\bibinfo {year} {2021})}\BibitemShut {NoStop}%
\bibitem [{\citenamefont {Alidoust}\ \emph {et~al.}(2015)\citenamefont
  {Alidoust}, \citenamefont {Halterman},\ and\ \citenamefont
  {Valls}}]{Klaus_2015}%
  \BibitemOpen
  \bibfield  {author} {\bibinfo {author} {\bibfnamefont {M.}~\bibnamefont
  {Alidoust}}, \bibinfo {author} {\bibfnamefont {K.}~\bibnamefont {Halterman}},
  \ and\ \bibinfo {author} {\bibfnamefont {O.~T.}\ \bibnamefont {Valls}},\
  }\href {\doibase 10.1103/PhysRevB.92.014508} {\bibfield  {journal} {\bibinfo
  {journal} {Phys. Rev. B}\ }\textbf {\bibinfo {volume} {92}},\ \bibinfo
  {pages} {014508} (\bibinfo {year} {2015})}\BibitemShut {NoStop}%
\bibitem [{\citenamefont {Zha}\ \emph {et~al.}(2015)\citenamefont {Zha},
  \citenamefont {Covaci}, \citenamefont {Peeters},\ and\ \citenamefont
  {Zhou}}]{Zha_2015}%
  \BibitemOpen
  \bibfield  {author} {\bibinfo {author} {\bibfnamefont {G.-Q.}\ \bibnamefont
  {Zha}}, \bibinfo {author} {\bibfnamefont {L.}~\bibnamefont {Covaci}},
  \bibinfo {author} {\bibfnamefont {F.~M.}\ \bibnamefont {Peeters}}, \ and\
  \bibinfo {author} {\bibfnamefont {S.-P.}\ \bibnamefont {Zhou}},\ }\href
  {\doibase 10.1103/PhysRevB.91.214504} {\bibfield  {journal} {\bibinfo
  {journal} {Phys. Rev. B}\ }\textbf {\bibinfo {volume} {91}},\ \bibinfo
  {pages} {214504} (\bibinfo {year} {2015})}\BibitemShut {NoStop}%
\bibitem [{\citenamefont {Wu}\ \emph {et~al.}(2012)\citenamefont {Wu},
  \citenamefont {Valls},\ and\ \citenamefont {Halterman}}]{Klaus_2012}%
  \BibitemOpen
  \bibfield  {author} {\bibinfo {author} {\bibfnamefont {C.-T.}\ \bibnamefont
  {Wu}}, \bibinfo {author} {\bibfnamefont {O.~T.}\ \bibnamefont {Valls}}, \
  and\ \bibinfo {author} {\bibfnamefont {K.}~\bibnamefont {Halterman}},\ }\href
  {\doibase 10.1103/PhysRevB.86.014523} {\bibfield  {journal} {\bibinfo
  {journal} {Phys. Rev. B}\ }\textbf {\bibinfo {volume} {86}},\ \bibinfo
  {pages} {014523} (\bibinfo {year} {2012})}\BibitemShut {NoStop}%
\bibitem [{\citenamefont {Halterman}\ \emph {et~al.}(2008)\citenamefont
  {Halterman}, \citenamefont {Valls},\ and\ \citenamefont
  {Barsic}}]{Klaus_2008}%
  \BibitemOpen
  \bibfield  {author} {\bibinfo {author} {\bibfnamefont {K.}~\bibnamefont
  {Halterman}}, \bibinfo {author} {\bibfnamefont {O.~T.}\ \bibnamefont
  {Valls}}, \ and\ \bibinfo {author} {\bibfnamefont {P.~H.}\ \bibnamefont
  {Barsic}},\ }\href {\doibase 10.1103/PhysRevB.77.174511} {\bibfield
  {journal} {\bibinfo  {journal} {Phys. Rev. B}\ }\textbf {\bibinfo {volume}
  {77}},\ \bibinfo {pages} {174511} (\bibinfo {year} {2008})}\BibitemShut
  {NoStop}%
\bibitem [{\citenamefont {Halterman}\ \emph {et~al.}(2007)\citenamefont
  {Halterman}, \citenamefont {Barsic},\ and\ \citenamefont
  {Valls}}]{Klaus_2007}%
  \BibitemOpen
  \bibfield  {author} {\bibinfo {author} {\bibfnamefont {K.}~\bibnamefont
  {Halterman}}, \bibinfo {author} {\bibfnamefont {P.~H.}\ \bibnamefont
  {Barsic}}, \ and\ \bibinfo {author} {\bibfnamefont {O.~T.}\ \bibnamefont
  {Valls}},\ }\href {\doibase 10.1103/PhysRevLett.99.127002} {\bibfield
  {journal} {\bibinfo  {journal} {Phys. Rev. Lett.}\ }\textbf {\bibinfo
  {volume} {99}},\ \bibinfo {pages} {127002} (\bibinfo {year}
  {2007})}\BibitemShut {NoStop}%
\bibitem [{\citenamefont {Fominov}\ \emph {et~al.}(2003)\citenamefont
  {Fominov}, \citenamefont {Golubov},\ and\ \citenamefont
  {Kupriyanov}}]{Fominov_2003}%
  \BibitemOpen
  \bibfield  {author} {\bibinfo {author} {\bibfnamefont {Y.~V.}\ \bibnamefont
  {Fominov}}, \bibinfo {author} {\bibfnamefont {A.~A.}\ \bibnamefont
  {Golubov}}, \ and\ \bibinfo {author} {\bibfnamefont {M.~Y.}\ \bibnamefont
  {Kupriyanov}},\ }\href {\doibase 10.1134/1.1591981} {\bibfield  {journal}
  {\bibinfo  {journal} {Journal of Experimental and Theoretical Physics
  Letters}\ }\textbf {\bibinfo {volume} {77}},\ \bibinfo {pages} {510}
  (\bibinfo {year} {2003})}\BibitemShut {NoStop}%
\bibitem [{\citenamefont {Hui}\ \emph {et~al.}(2015)\citenamefont {Hui},
  \citenamefont {Brydon}, \citenamefont {Sau}, \citenamefont {Tewari},\ and\
  \citenamefont {Sarma}}]{Hui_2015}%
  \BibitemOpen
  \bibfield  {author} {\bibinfo {author} {\bibfnamefont {H.-Y.}\ \bibnamefont
  {Hui}}, \bibinfo {author} {\bibfnamefont {P.~M.~R.}\ \bibnamefont {Brydon}},
  \bibinfo {author} {\bibfnamefont {J.~D.}\ \bibnamefont {Sau}}, \bibinfo
  {author} {\bibfnamefont {S.}~\bibnamefont {Tewari}}, \ and\ \bibinfo {author}
  {\bibfnamefont {S.~D.}\ \bibnamefont {Sarma}},\ }\href {\doibase
  10.1038/srep08880} {\bibfield  {journal} {\bibinfo  {journal} {Scientific
  Reports}\ }\textbf {\bibinfo {volume} {5}} (\bibinfo {year} {2015}),\
  10.1038/srep08880}\BibitemShut {NoStop}%
\bibitem [{\citenamefont {Halterman}\ and\ \citenamefont
  {Alidoust}(2018)}]{Klaus_2018_SF}%
  \BibitemOpen
  \bibfield  {author} {\bibinfo {author} {\bibfnamefont {K.}~\bibnamefont
  {Halterman}}\ and\ \bibinfo {author} {\bibfnamefont {M.}~\bibnamefont
  {Alidoust}},\ }\href {\doibase 10.1103/PhysRevB.98.134510} {\bibfield
  {journal} {\bibinfo  {journal} {Phys. Rev. B}\ }\textbf {\bibinfo {volume}
  {98}},\ \bibinfo {pages} {134510} (\bibinfo {year} {2018})}\BibitemShut
  {NoStop}%
\bibitem [{\citenamefont {Halterman}\ and\ \citenamefont
  {Valls}(2005)}]{Klaus_2005}%
  \BibitemOpen
  \bibfield  {author} {\bibinfo {author} {\bibfnamefont {K.}~\bibnamefont
  {Halterman}}\ and\ \bibinfo {author} {\bibfnamefont {O.~T.}\ \bibnamefont
  {Valls}},\ }\href {\doibase 10.1103/PhysRevB.72.060514} {\bibfield  {journal}
  {\bibinfo  {journal} {Phys. Rev. B}\ }\textbf {\bibinfo {volume} {72}},\
  \bibinfo {pages} {060514(R)} (\bibinfo {year} {2005})}\BibitemShut {NoStop}%
\bibitem [{\citenamefont {Halterman}\ and\ \citenamefont
  {Valls}(2003)}]{Klaus_2003}%
  \BibitemOpen
  \bibfield  {author} {\bibinfo {author} {\bibfnamefont {K.}~\bibnamefont
  {Halterman}}\ and\ \bibinfo {author} {\bibfnamefont {O.~T.}\ \bibnamefont
  {Valls}},\ }\href {\doibase https://doi.org/10.1016/S0921-4534(03)01095-5}
  {\bibfield  {journal} {\bibinfo  {journal} {Physica C: Superconductivity}\
  }\textbf {\bibinfo {volume} {397}},\ \bibinfo {pages} {151} (\bibinfo {year}
  {2003})}\BibitemShut {NoStop}%
\bibitem [{\citenamefont {Bergeret}\ \emph {et~al.}(2001)\citenamefont
  {Bergeret}, \citenamefont {Volkov},\ and\ \citenamefont
  {Efetov}}]{Bergeret_2001}%
  \BibitemOpen
  \bibfield  {author} {\bibinfo {author} {\bibfnamefont {F.~S.}\ \bibnamefont
  {Bergeret}}, \bibinfo {author} {\bibfnamefont {A.~F.}\ \bibnamefont
  {Volkov}}, \ and\ \bibinfo {author} {\bibfnamefont {K.~B.}\ \bibnamefont
  {Efetov}},\ }\href {\doibase 10.1103/PhysRevLett.86.4096} {\bibfield
  {journal} {\bibinfo  {journal} {Phys. Rev. Lett.}\ }\textbf {\bibinfo
  {volume} {86}},\ \bibinfo {pages} {4096} (\bibinfo {year}
  {2001})}\BibitemShut {NoStop}%
\bibitem [{\citenamefont {Halterman}\ and\ \citenamefont
  {Valls}(2001)}]{Klaus_2001}%
  \BibitemOpen
  \bibfield  {author} {\bibinfo {author} {\bibfnamefont {K.}~\bibnamefont
  {Halterman}}\ and\ \bibinfo {author} {\bibfnamefont {O.~T.}\ \bibnamefont
  {Valls}},\ }\href {\doibase 10.1103/PhysRevB.65.014509} {\bibfield  {journal}
  {\bibinfo  {journal} {Phys. Rev. B}\ }\textbf {\bibinfo {volume} {65}},\
  \bibinfo {pages} {014509} (\bibinfo {year} {2001})}\BibitemShut {NoStop}%
\bibitem [{\citenamefont {Yang}\ \emph {et~al.}(2021)\citenamefont {Yang},
  \citenamefont {Ciccarelli},\ and\ \citenamefont
  {Robinson}}]{SCspintronics_21}%
  \BibitemOpen
  \bibfield  {author} {\bibinfo {author} {\bibfnamefont {G.}~\bibnamefont
  {Yang}}, \bibinfo {author} {\bibfnamefont {C.}~\bibnamefont {Ciccarelli}}, \
  and\ \bibinfo {author} {\bibfnamefont {J.~W.~A.}\ \bibnamefont {Robinson}},\
  }\href {\doibase 10.1063/5.0048904} {\bibfield  {journal} {\bibinfo
  {journal} {APL Materials}\ }\textbf {\bibinfo {volume} {9}},\ \bibinfo
  {pages} {050703} (\bibinfo {year} {2021})},\ \Eprint
  {http://arxiv.org/abs/https://doi.org/10.1063/5.0048904}
  {https://doi.org/10.1063/5.0048904} \BibitemShut {NoStop}%
\bibitem [{\citenamefont {Proshin}\ \emph {et~al.}(2002)\citenamefont
  {Proshin}, \citenamefont {Izyumov},\ and\ \citenamefont
  {Khusainov}}]{Khusainov_02}%
  \BibitemOpen
  \bibfield  {author} {\bibinfo {author} {\bibfnamefont {Y.~N.}\ \bibnamefont
  {Proshin}}, \bibinfo {author} {\bibfnamefont {Y.~A.}\ \bibnamefont
  {Izyumov}}, \ and\ \bibinfo {author} {\bibfnamefont {M.~G.}\ \bibnamefont
  {Khusainov}},\ }\href {\doibase
  https://doi.org/10.1016/S0921-4534(01)01009-7} {\bibfield  {journal}
  {\bibinfo  {journal} {Physica C: Superconductivity}\ }\textbf {\bibinfo
  {volume} {367}},\ \bibinfo {pages} {181} (\bibinfo {year}
  {2002})}\BibitemShut {NoStop}%
\bibitem [{\citenamefont {Halterman}\ and\ \citenamefont
  {Alidoust}(2016)}]{Klaus_2016}%
  \BibitemOpen
  \bibfield  {author} {\bibinfo {author} {\bibfnamefont {K.}~\bibnamefont
  {Halterman}}\ and\ \bibinfo {author} {\bibfnamefont {M.}~\bibnamefont
  {Alidoust}},\ }\href {\doibase 10.1103/PhysRevB.94.064503} {\bibfield
  {journal} {\bibinfo  {journal} {Phys. Rev. B}\ }\textbf {\bibinfo {volume}
  {94}},\ \bibinfo {pages} {064503} (\bibinfo {year} {2016})}\BibitemShut
  {NoStop}%
\bibitem [{\citenamefont {Tagirov}(1999)}]{SpinSwitch}%
  \BibitemOpen
  \bibfield  {author} {\bibinfo {author} {\bibfnamefont {L.~R.}\ \bibnamefont
  {Tagirov}},\ }\href {\doibase 10.1103/PhysRevLett.83.2058} {\bibfield
  {journal} {\bibinfo  {journal} {Phys. Rev. Lett.}\ }\textbf {\bibinfo
  {volume} {83}},\ \bibinfo {pages} {2058} (\bibinfo {year}
  {1999})}\BibitemShut {NoStop}%
\bibitem [{\citenamefont {Alidoust}\ and\ \citenamefont
  {Halterman}(2018)}]{Klaus_2018_spinvalve}%
  \BibitemOpen
  \bibfield  {author} {\bibinfo {author} {\bibfnamefont {M.}~\bibnamefont
  {Alidoust}}\ and\ \bibinfo {author} {\bibfnamefont {K.}~\bibnamefont
  {Halterman}},\ }\href {\doibase 10.1103/PhysRevB.97.064517} {\bibfield
  {journal} {\bibinfo  {journal} {Phys. Rev. B}\ }\textbf {\bibinfo {volume}
  {97}},\ \bibinfo {pages} {064517} (\bibinfo {year} {2018})}\BibitemShut
  {NoStop}%
\bibitem [{\citenamefont {Buzdin}(2005)}]{Budzin_05}%
  \BibitemOpen
  \bibfield  {author} {\bibinfo {author} {\bibfnamefont {A.~I.}\ \bibnamefont
  {Buzdin}},\ }\href {\doibase 10.1103/RevModPhys.77.935} {\bibfield  {journal}
  {\bibinfo  {journal} {Rev. Mod. Phys.}\ }\textbf {\bibinfo {volume} {77}},\
  \bibinfo {pages} {935} (\bibinfo {year} {2005})}\BibitemShut {NoStop}%
\bibitem [{\citenamefont {Anderson}\ \emph {et~al.}(1999)\citenamefont
  {Anderson}, \citenamefont {Bai}, \citenamefont {Bischof}, \citenamefont
  {Blackford}, \citenamefont {Demmel}, \citenamefont {Dongarra}, \citenamefont
  {Du~Croz}, \citenamefont {Greenbaum}, \citenamefont {Hammarling},
  \citenamefont {McKenney},\ and\ \citenamefont {Sorensen}}]{lapack99}%
  \BibitemOpen
  \bibfield  {author} {\bibinfo {author} {\bibfnamefont {E.}~\bibnamefont
  {Anderson}}, \bibinfo {author} {\bibfnamefont {Z.}~\bibnamefont {Bai}},
  \bibinfo {author} {\bibfnamefont {C.}~\bibnamefont {Bischof}}, \bibinfo
  {author} {\bibfnamefont {S.}~\bibnamefont {Blackford}}, \bibinfo {author}
  {\bibfnamefont {J.}~\bibnamefont {Demmel}}, \bibinfo {author} {\bibfnamefont
  {J.}~\bibnamefont {Dongarra}}, \bibinfo {author} {\bibfnamefont
  {J.}~\bibnamefont {Du~Croz}}, \bibinfo {author} {\bibfnamefont
  {A.}~\bibnamefont {Greenbaum}}, \bibinfo {author} {\bibfnamefont
  {S.}~\bibnamefont {Hammarling}}, \bibinfo {author} {\bibfnamefont
  {A.}~\bibnamefont {McKenney}}, \ and\ \bibinfo {author} {\bibfnamefont
  {D.}~\bibnamefont {Sorensen}},\ }\href@noop {} {\emph {\bibinfo {title}
  {{LAPACK} Users' Guide}}},\ \bibinfo {edition} {3rd}\ ed.\ (\bibinfo
  {publisher} {Society for Industrial and Applied Mathematics},\ \bibinfo
  {address} {Philadelphia, PA},\ \bibinfo {year} {1999})\BibitemShut {NoStop}%
\bibitem [{\citenamefont {Arabas}\ \emph {et~al.}(2014)\citenamefont {Arabas},
  \citenamefont {Jarecka}, \citenamefont {Jaruga},\ and\ \citenamefont
  {Fija{\l}kowski}}]{FastFortran}%
  \BibitemOpen
  \bibfield  {author} {\bibinfo {author} {\bibfnamefont {S.}~\bibnamefont
  {Arabas}}, \bibinfo {author} {\bibfnamefont {D.}~\bibnamefont {Jarecka}},
  \bibinfo {author} {\bibfnamefont {A.}~\bibnamefont {Jaruga}}, \ and\ \bibinfo
  {author} {\bibfnamefont {M.}~\bibnamefont {Fija{\l}kowski}},\ }\href@noop {}
  {\bibfield  {journal} {\bibinfo  {journal} {Scientific Programming}\ }\textbf
  {\bibinfo {volume} {22}},\ \bibinfo {pages} {201} (\bibinfo {year}
  {2014})}\BibitemShut {NoStop}%
\bibitem [{\citenamefont {Zhu}(2016)}]{Jianxin}%
  \BibitemOpen
  \bibfield  {author} {\bibinfo {author} {\bibfnamefont {J.-X.}\ \bibnamefont
  {Zhu}},\ }\href@noop {} {\emph {\bibinfo {title} {Bogoliubov de Gennes
  Methods and its Applications}}}\ (\bibinfo  {publisher} {Springer},\ \bibinfo
  {year} {2016})\BibitemShut {NoStop}%
\bibitem [{\citenamefont {Bogoliubov}(1958)}]{Bogoliubov:1958km}%
  \BibitemOpen
  \bibfield  {author} {\bibinfo {author} {\bibfnamefont {N.~N.}\ \bibnamefont
  {Bogoliubov}},\ }\href {\doibase 10.1007/BF02745585} {\bibfield  {journal}
  {\bibinfo  {journal} {Nuovo Cim.}\ }\textbf {\bibinfo {volume} {7}},\
  \bibinfo {pages} {794} (\bibinfo {year} {1958})}\BibitemShut {NoStop}%
\bibitem [{\citenamefont {Reeg}\ and\ \citenamefont
  {Maslov}(2016)}]{Reeg_2016}%
  \BibitemOpen
  \bibfield  {author} {\bibinfo {author} {\bibfnamefont {C.~R.}\ \bibnamefont
  {Reeg}}\ and\ \bibinfo {author} {\bibfnamefont {D.~L.}\ \bibnamefont
  {Maslov}},\ }\href {\doibase 10.1103/PhysRevB.94.020501} {\bibfield
  {journal} {\bibinfo  {journal} {Phys. Rev. B}\ }\textbf {\bibinfo {volume}
  {94}},\ \bibinfo {pages} {020501(R)} (\bibinfo {year} {2016})}\BibitemShut
  {NoStop}%
\bibitem [{\citenamefont {Beenakker}(2005)}]{Beenakker_2005}%
  \BibitemOpen
  \bibfield  {author} {\bibinfo {author} {\bibfnamefont {C.}~\bibnamefont
  {Beenakker}},\ }in\ \href {\doibase 10.1007/11358817_4} {\emph {\bibinfo
  {booktitle} {Quantum Dots: a Doorway to Nanoscale Physics}}}\ (\bibinfo
  {publisher} {Springer Berlin Heidelberg},\ \bibinfo {year} {2005})\ pp.\
  \bibinfo {pages} {131--174}\BibitemShut {NoStop}%
\bibitem [{\citenamefont {Vaitiekenas}\ \emph {et~al.}(2020)\citenamefont
  {Vaitiekenas}, \citenamefont {Winkler}, \citenamefont {van Heck},
  \citenamefont {Karzig}, \citenamefont {Deng}, \citenamefont {Flensberg},
  \citenamefont {Glazman}, \citenamefont {Nayak}, \citenamefont {Krogstrup},
  \citenamefont {Lutchyn},\ and\ \citenamefont
  {Marcus}}]{FluxInducedMajorana2018}%
  \BibitemOpen
  \bibfield  {author} {\bibinfo {author} {\bibfnamefont {S.}~\bibnamefont
  {Vaitiekenas}}, \bibinfo {author} {\bibfnamefont {G.~W.}\ \bibnamefont
  {Winkler}}, \bibinfo {author} {\bibfnamefont {B.}~\bibnamefont {van Heck}},
  \bibinfo {author} {\bibfnamefont {T.}~\bibnamefont {Karzig}}, \bibinfo
  {author} {\bibfnamefont {M.-T.}\ \bibnamefont {Deng}}, \bibinfo {author}
  {\bibfnamefont {K.}~\bibnamefont {Flensberg}}, \bibinfo {author}
  {\bibfnamefont {L.~I.}\ \bibnamefont {Glazman}}, \bibinfo {author}
  {\bibfnamefont {C.}~\bibnamefont {Nayak}}, \bibinfo {author} {\bibfnamefont
  {P.}~\bibnamefont {Krogstrup}}, \bibinfo {author} {\bibfnamefont {R.~M.}\
  \bibnamefont {Lutchyn}}, \ and\ \bibinfo {author} {\bibfnamefont {C.~M.}\
  \bibnamefont {Marcus}},\ }\href {\doibase 10.1126/science.aav3392} {\bibfield
   {journal} {\bibinfo  {journal} {Science}\ }\textbf {\bibinfo {volume} {367}}
  (\bibinfo {year} {2020}),\ 10.1126/science.aav3392}\BibitemShut {NoStop}%
\bibitem [{\citenamefont {{Pe{\~n}aranda}}\ \emph {et~al.}(2019)\citenamefont
  {{Pe{\~n}aranda}}, \citenamefont {{Aguado}}, \citenamefont {{San-Jose}},\
  and\ \citenamefont {{Prada}}}]{evenodd}%
  \BibitemOpen
  \bibfield  {author} {\bibinfo {author} {\bibfnamefont {F.}~\bibnamefont
  {{Pe{\~n}aranda}}}, \bibinfo {author} {\bibfnamefont {R.}~\bibnamefont
  {{Aguado}}}, \bibinfo {author} {\bibfnamefont {P.}~\bibnamefont
  {{San-Jose}}}, \ and\ \bibinfo {author} {\bibfnamefont {E.}~\bibnamefont
  {{Prada}}},\ }\href@noop {} {\bibfield  {journal} {\bibinfo  {journal} {arXiv
  e-prints}\ ,\ \bibinfo {eid} {arXiv:1911.06805}} (\bibinfo {year} {2019})},\
  \Eprint {http://arxiv.org/abs/1911.06805} {arXiv:1911.06805
  [cond-mat.mes-hall]} \BibitemShut {NoStop}%
\bibitem [{\citenamefont {Kringh\o{}j}\ \emph {et~al.}(2021)\citenamefont
  {Kringh\o{}j}, \citenamefont {Winkler}, \citenamefont {Larsen}, \citenamefont
  {Sabonis}, \citenamefont {Erlandsson}, \citenamefont {Krogstrup},
  \citenamefont {van Heck}, \citenamefont {Petersson},\ and\ \citenamefont
  {Marcus}}]{Marcus21}%
  \BibitemOpen
  \bibfield  {author} {\bibinfo {author} {\bibfnamefont {A.}~\bibnamefont
  {Kringh\o{}j}}, \bibinfo {author} {\bibfnamefont {G.~W.}\ \bibnamefont
  {Winkler}}, \bibinfo {author} {\bibfnamefont {T.~W.}\ \bibnamefont {Larsen}},
  \bibinfo {author} {\bibfnamefont {D.}~\bibnamefont {Sabonis}}, \bibinfo
  {author} {\bibfnamefont {O.}~\bibnamefont {Erlandsson}}, \bibinfo {author}
  {\bibfnamefont {P.}~\bibnamefont {Krogstrup}}, \bibinfo {author}
  {\bibfnamefont {B.}~\bibnamefont {van Heck}}, \bibinfo {author}
  {\bibfnamefont {K.~D.}\ \bibnamefont {Petersson}}, \ and\ \bibinfo {author}
  {\bibfnamefont {C.~M.}\ \bibnamefont {Marcus}},\ }\href {\doibase
  10.1103/PhysRevLett.126.047701} {\bibfield  {journal} {\bibinfo  {journal}
  {Phys. Rev. Lett.}\ }\textbf {\bibinfo {volume} {126}},\ \bibinfo {pages}
  {047701} (\bibinfo {year} {2021})}\BibitemShut {NoStop}%
\bibitem [{\citenamefont {Stanescu}\ and\ \citenamefont
  {Das~Sarma}(2022)}]{Stanescu_2022}%
  \BibitemOpen
  \bibfield  {author} {\bibinfo {author} {\bibfnamefont {T.~D.}\ \bibnamefont
  {Stanescu}}\ and\ \bibinfo {author} {\bibfnamefont {S.}~\bibnamefont
  {Das~Sarma}},\ }\href {\doibase 10.1103/PhysRevB.106.085429} {\bibfield
  {journal} {\bibinfo  {journal} {Phys. Rev. B}\ }\textbf {\bibinfo {volume}
  {106}},\ \bibinfo {pages} {085429} (\bibinfo {year} {2022})}\BibitemShut
  {NoStop}%
\bibitem [{\citenamefont {Kiendl}\ \emph {et~al.}(2019)\citenamefont {Kiendl},
  \citenamefont {von Oppen},\ and\ \citenamefont {Brouwer}}]{Kiendl_2019}%
  \BibitemOpen
  \bibfield  {author} {\bibinfo {author} {\bibfnamefont {T.}~\bibnamefont
  {Kiendl}}, \bibinfo {author} {\bibfnamefont {F.}~\bibnamefont {von Oppen}}, \
  and\ \bibinfo {author} {\bibfnamefont {P.~W.}\ \bibnamefont {Brouwer}},\
  }\href {\doibase 10.1103/PhysRevB.100.035426} {\bibfield  {journal} {\bibinfo
   {journal} {Phys. Rev. B}\ }\textbf {\bibinfo {volume} {100}},\ \bibinfo
  {pages} {035426} (\bibinfo {year} {2019})}\BibitemShut {NoStop}%
\bibitem [{\citenamefont {Zhang}\ \emph {et~al.}(2019)\citenamefont {Zhang},
  \citenamefont {Liu}, \citenamefont {Wimmer},\ and\ \citenamefont
  {Kouwenhoven}}]{Zhang2019}%
  \BibitemOpen
  \bibfield  {author} {\bibinfo {author} {\bibfnamefont {H.}~\bibnamefont
  {Zhang}}, \bibinfo {author} {\bibfnamefont {D.~E.}\ \bibnamefont {Liu}},
  \bibinfo {author} {\bibfnamefont {M.}~\bibnamefont {Wimmer}}, \ and\ \bibinfo
  {author} {\bibfnamefont {L.~P.}\ \bibnamefont {Kouwenhoven}},\ }\href
  {\doibase 10.1038/s41467-019-13133-1} {\bibfield  {journal} {\bibinfo
  {journal} {Nature Communications}\ }\textbf {\bibinfo {volume} {10}},\
  \bibinfo {pages} {5128} (\bibinfo {year} {2019})}\BibitemShut {NoStop}%
\bibitem [{\citenamefont {Maier}\ \emph {et~al.}(2014)\citenamefont {Maier},
  \citenamefont {Klinovaja},\ and\ \citenamefont {Loss}}]{mfcs2014}%
  \BibitemOpen
  \bibfield  {author} {\bibinfo {author} {\bibfnamefont {F.}~\bibnamefont
  {Maier}}, \bibinfo {author} {\bibfnamefont {J.}~\bibnamefont {Klinovaja}}, \
  and\ \bibinfo {author} {\bibfnamefont {D.}~\bibnamefont {Loss}},\ }\href
  {\doibase 10.1103/PhysRevB.90.195421} {\bibfield  {journal} {\bibinfo
  {journal} {Phys. Rev. B}\ }\textbf {\bibinfo {volume} {90}},\ \bibinfo
  {pages} {195421} (\bibinfo {year} {2014})}\BibitemShut {NoStop}%
\bibitem [{\citenamefont {Stanescu}\ and\ \citenamefont
  {Tewari}(2013)}]{Stanescu2013}%
  \BibitemOpen
  \bibfield  {author} {\bibinfo {author} {\bibfnamefont {T.~D.}\ \bibnamefont
  {Stanescu}}\ and\ \bibinfo {author} {\bibfnamefont {S.}~\bibnamefont
  {Tewari}},\ }\href@noop {} {\bibfield  {journal} {\bibinfo  {journal}
  {Journal of Physics: Condensed Matter}\ }\textbf {\bibinfo {volume} {25}},\
  \bibinfo {pages} {233201} (\bibinfo {year} {2013})}\BibitemShut {NoStop}%
\bibitem [{\citenamefont {G{\"u}l}\ \emph {et~al.}(2018)\citenamefont
  {G{\"u}l}, \citenamefont {Zhang}, \citenamefont {Bommer}, \citenamefont
  {de~Moor}, \citenamefont {Car}, \citenamefont {Plissard}, \citenamefont
  {Bakkers}, \citenamefont {Geresdi}, \citenamefont {Watanabe}, \citenamefont
  {Taniguchi},\ and\ \citenamefont {Kouwenhoven}}]{Gul2018}%
  \BibitemOpen
  \bibfield  {author} {\bibinfo {author} {\bibfnamefont {{\"O}.}~\bibnamefont
  {G{\"u}l}}, \bibinfo {author} {\bibfnamefont {H.}~\bibnamefont {Zhang}},
  \bibinfo {author} {\bibfnamefont {J.~D.~S.}\ \bibnamefont {Bommer}}, \bibinfo
  {author} {\bibfnamefont {M.~W.~A.}\ \bibnamefont {de~Moor}}, \bibinfo
  {author} {\bibfnamefont {D.}~\bibnamefont {Car}}, \bibinfo {author}
  {\bibfnamefont {S.~R.}\ \bibnamefont {Plissard}}, \bibinfo {author}
  {\bibfnamefont {E.~P. A.~M.}\ \bibnamefont {Bakkers}}, \bibinfo {author}
  {\bibfnamefont {A.}~\bibnamefont {Geresdi}}, \bibinfo {author} {\bibfnamefont
  {K.}~\bibnamefont {Watanabe}}, \bibinfo {author} {\bibfnamefont
  {T.}~\bibnamefont {Taniguchi}}, \ and\ \bibinfo {author} {\bibfnamefont
  {L.~P.}\ \bibnamefont {Kouwenhoven}},\ }\href {\doibase
  10.1038/s41565-017-0032-8} {\bibfield  {journal} {\bibinfo  {journal} {Nature
  Nanotechnology}\ }\textbf {\bibinfo {volume} {13}},\ \bibinfo {pages} {192}
  (\bibinfo {year} {2018})}\BibitemShut {NoStop}%
\bibitem [{\citenamefont {Millo}\ and\ \citenamefont
  {Koren}(2018)}]{Millo_etal}%
  \BibitemOpen
  \bibfield  {author} {\bibinfo {author} {\bibfnamefont {O.}~\bibnamefont
  {Millo}}\ and\ \bibinfo {author} {\bibfnamefont {G.}~\bibnamefont {Koren}},\
  }\href@noop {} {\bibfield  {journal} {\bibinfo  {journal} {Philos. Trans.
  Royal Soc. A}\ }\textbf {\bibinfo {volume} {376}},\ \bibinfo {pages}
  {20140143} (\bibinfo {year} {2018})}\BibitemShut {NoStop}%
\bibitem [{\citenamefont {Ko}\ \emph {et~al.}(2022)\citenamefont {Ko},
  \citenamefont {Lado},\ and\ \citenamefont {Maksymovych}}]{Ko2022}%
  \BibitemOpen
  \bibfield  {author} {\bibinfo {author} {\bibfnamefont {W.}~\bibnamefont
  {Ko}}, \bibinfo {author} {\bibfnamefont {J.~L.}\ \bibnamefont {Lado}}, \ and\
  \bibinfo {author} {\bibfnamefont {P.}~\bibnamefont {Maksymovych}},\ }\href
  {\doibase 10.1021/acs.nanolett.2c00697} {\bibfield  {journal} {\bibinfo
  {journal} {Nano Letters}\ }\textbf {\bibinfo {volume} {22}},\ \bibinfo
  {pages} {4042} (\bibinfo {year} {2022})}\BibitemShut {NoStop}%
\bibitem [{\citenamefont {Hashisaka}\ \emph {et~al.}(2021)\citenamefont
  {Hashisaka}, \citenamefont {Jonckheere}, \citenamefont {Akiho}, \citenamefont
  {Sasaki}, \citenamefont {Rech}, \citenamefont {Martin},\ and\ \citenamefont
  {Muraki}}]{Hashisaka2021}%
  \BibitemOpen
  \bibfield  {author} {\bibinfo {author} {\bibfnamefont {M.}~\bibnamefont
  {Hashisaka}}, \bibinfo {author} {\bibfnamefont {T.}~\bibnamefont
  {Jonckheere}}, \bibinfo {author} {\bibfnamefont {T.}~\bibnamefont {Akiho}},
  \bibinfo {author} {\bibfnamefont {S.}~\bibnamefont {Sasaki}}, \bibinfo
  {author} {\bibfnamefont {J.}~\bibnamefont {Rech}}, \bibinfo {author}
  {\bibfnamefont {T.}~\bibnamefont {Martin}}, \ and\ \bibinfo {author}
  {\bibfnamefont {K.}~\bibnamefont {Muraki}},\ }\href {\doibase
  10.1038/s41467-021-23160-6} {\bibfield  {journal} {\bibinfo  {journal}
  {Nature Communications}\ }\textbf {\bibinfo {volume} {12}},\ \bibinfo {pages}
  {2794} (\bibinfo {year} {2021})}\BibitemShut {NoStop}%
\bibitem [{\citenamefont {Chalsani}\ \emph {et~al.}(2007)\citenamefont
  {Chalsani}, \citenamefont {Upadhyay}, \citenamefont {Ozatay},\ and\
  \citenamefont {Buhrman}}]{Chalsani_2007}%
  \BibitemOpen
  \bibfield  {author} {\bibinfo {author} {\bibfnamefont {P.}~\bibnamefont
  {Chalsani}}, \bibinfo {author} {\bibfnamefont {S.~K.}\ \bibnamefont
  {Upadhyay}}, \bibinfo {author} {\bibfnamefont {O.}~\bibnamefont {Ozatay}}, \
  and\ \bibinfo {author} {\bibfnamefont {R.~A.}\ \bibnamefont {Buhrman}},\
  }\href {\doibase 10.1103/PhysRevB.75.094417} {\bibfield  {journal} {\bibinfo
  {journal} {Phys. Rev. B}\ }\textbf {\bibinfo {volume} {75}},\ \bibinfo
  {pages} {094417} (\bibinfo {year} {2007})}\BibitemShut {NoStop}%
\bibitem [{\citenamefont {Gupta}\ \emph {et~al.}(2006)\citenamefont {Gupta},
  \citenamefont {Crétinon}, \citenamefont {Pannetier},\ and\ \citenamefont
  {Courtois}}]{Gupta_2005}%
  \BibitemOpen
  \bibfield  {author} {\bibinfo {author} {\bibfnamefont {A.}~\bibnamefont
  {Gupta}}, \bibinfo {author} {\bibfnamefont {L.}~\bibnamefont {Crétinon}},
  \bibinfo {author} {\bibfnamefont {B.}~\bibnamefont {Pannetier}}, \ and\
  \bibinfo {author} {\bibfnamefont {H.}~\bibnamefont {Courtois}},\ }\href
  {\doibase 10.1007/BF02704953} {\bibfield  {journal} {\bibinfo  {journal}
  {Pramana - Journal of Physics}\ }\textbf {\bibinfo {volume} {66}},\ \bibinfo
  {pages} {251} (\bibinfo {year} {2006})}\BibitemShut {NoStop}%
\bibitem [{\citenamefont {Mortensen}\ \emph {et~al.}(1999)\citenamefont
  {Mortensen}, \citenamefont {Flensberg},\ and\ \citenamefont
  {Jauho}}]{Mortensen_1999}%
  \BibitemOpen
  \bibfield  {author} {\bibinfo {author} {\bibfnamefont {N.~A.}\ \bibnamefont
  {Mortensen}}, \bibinfo {author} {\bibfnamefont {K.}~\bibnamefont
  {Flensberg}}, \ and\ \bibinfo {author} {\bibfnamefont {A.-P.}\ \bibnamefont
  {Jauho}},\ }\href {\doibase 10.1103/physrevb.59.10176} {\bibfield  {journal}
  {\bibinfo  {journal} {Physical Review B}\ }\textbf {\bibinfo {volume} {59}},\
  \bibinfo {pages} {10176} (\bibinfo {year} {1999})}\BibitemShut {NoStop}%
\bibitem [{\citenamefont {Sipr}\ and\ \citenamefont {{L.
  Gy{\"o}rffy}}(1997)}]{Sipr_1997}%
  \BibitemOpen
  \bibfield  {author} {\bibinfo {author} {\bibfnamefont {O.}~\bibnamefont
  {Sipr}}\ and\ \bibinfo {author} {\bibfnamefont {B.}~\bibnamefont {{L.
  Gy{\"o}rffy}}},\ }\href {\doibase 10.1007/BF02399632} {\bibfield  {journal}
  {\bibinfo  {journal} {Journal of Low Temperature Physics}\ }\textbf {\bibinfo
  {volume} {106}},\ \bibinfo {pages} {315} (\bibinfo {year}
  {1997})}\BibitemShut {NoStop}%
\bibitem [{\citenamefont {Beenakker}(1997)}]{Beenakker_1997}%
  \BibitemOpen
  \bibfield  {author} {\bibinfo {author} {\bibfnamefont {C.~W.~J.}\
  \bibnamefont {Beenakker}},\ }\href {\doibase 10.1103/RevModPhys.69.731}
  {\bibfield  {journal} {\bibinfo  {journal} {Rev. Mod. Phys.}\ }\textbf
  {\bibinfo {volume} {69}},\ \bibinfo {pages} {731} (\bibinfo {year}
  {1997})}\BibitemShut {NoStop}%
\bibitem [{\citenamefont {B\'eri}\ \emph {et~al.}(2009)\citenamefont {B\'eri},
  \citenamefont {Kupferschmidt}, \citenamefont {Beenakker},\ and\ \citenamefont
  {Brouwer}}]{Beenakker_2009}%
  \BibitemOpen
  \bibfield  {author} {\bibinfo {author} {\bibfnamefont {B.}~\bibnamefont
  {B\'eri}}, \bibinfo {author} {\bibfnamefont {J.~N.}\ \bibnamefont
  {Kupferschmidt}}, \bibinfo {author} {\bibfnamefont {C.~W.~J.}\ \bibnamefont
  {Beenakker}}, \ and\ \bibinfo {author} {\bibfnamefont {P.~W.}\ \bibnamefont
  {Brouwer}},\ }\href {\doibase 10.1103/PhysRevB.79.024517} {\bibfield
  {journal} {\bibinfo  {journal} {Phys. Rev. B}\ }\textbf {\bibinfo {volume}
  {79}},\ \bibinfo {pages} {024517} (\bibinfo {year} {2009})}\BibitemShut
  {NoStop}%
\bibitem [{\citenamefont {Stanescu}\ and\ \citenamefont {{Das
  Sarma}}(2017)}]{TudorSarma_2017}%
  \BibitemOpen
  \bibfield  {author} {\bibinfo {author} {\bibfnamefont {T.~D.}\ \bibnamefont
  {Stanescu}}\ and\ \bibinfo {author} {\bibfnamefont {S.}~\bibnamefont {{Das
  Sarma}}},\ }\href {\doibase 10.1103/PhysRevB.96.014510} {\bibfield  {journal}
  {\bibinfo  {journal} {Phys. Rev. B}\ }\textbf {\bibinfo {volume} {96}},\
  \bibinfo {pages} {014510} (\bibinfo {year} {2017})}\BibitemShut {NoStop}%
\end{thebibliography}%
\end{document}